\documentclass{aa}
\usepackage{float}
\usepackage{graphicx}
\usepackage{txfonts}
\usepackage{array}
\usepackage{enumitem}
\usepackage{amsmath}	
\usepackage{subfig} 
\usepackage{color}
\usepackage{url} 
\newcommand{\SigmaRe}{$\Sigma_{R_\mathrm{e}}$}

\begin{document}
   \title{IllustrisTNG Insights: Factors Affecting the Presence of Bars in Disk Galaxies}

   \author{Shuai Lu
          \inst{1}
          \and
          Min Du\inst{1}\fnmsep\thanks{Any response should be directed to E-mail: dumin@xmu.edu.cn}
          \and
          Victor P. Debattista\inst{2}
          }

   \institute{Department of Astronomy, Xiamen University, Xiamen, Fujian 361005, P.R. China\\
         \and
             Jeremiah Horrocks Institute, University of Central Lancashire, Preston, PR1 2HE, UK\\
             }


  \abstract
   {}
   {Bars are important in the secular evolution of galaxies. The reproduction of the fraction and size of bars can also be regarded as an indicator for reproducing correct internal dynamical processes, which is a crucial test for cosmological simulations. This study is aimed at exploring the reasons why some galaxies have bars at redshift $z=0$ while others do not.}
   {We use ellipse fitting to measure the properties and evolution of bars in the IllustrisTNG cosmological simulation. By using the K-S two-sample test and tracing their evolutionary changes, we analyze the parameter differences between barred and unbarred galaxies. The properties of galaxies with short bars are also studied.}
   {When tracing all disk galaxies at \(z = 0\) back to \(z=1\), all of them show similar bar features at $z=1$. The fraction of bars increases in barred and short-bar galaxies but decreases in unbarred galaxies during \(z = 1-0\). In the case of disk galaxies with stellar mass \(\log(M_*/M_\odot)> 10.8\), nurture (mainly mergers) plays the most important role in suppressing or destroying bars. Bars are more likely to endure in galaxies that experience fewer mergers, which can be quantified by smaller stellar halos and ex-situ mass fractions. Approximately 60\% of the unbarred galaxies in the local Universe once had a bar. In contrast, the lack of responsiveness to bar instabilities (resulting in a larger Toomre-\(Q\) parameter) due to a less compact nature plays an important role in generating unbarred disk galaxies with stellar mass \(\log(M_*/M_\odot)<10.8\). Moreover, short bars generally form at a similar time to normal bars, during which they either grow mildly or contract significantly. The fact that IllustrisTNG simulations produce too many galaxies with short bars indicates that the dynamical properties of the central regions in IllustrisTNG galaxies are less affected by external factors, such as mergers and gas inflows.}
    {}


\maketitle


\mathchardef\mhyphen="2D
\section{Introduction}
Bars, which are linear structures at the centers of disk galaxies, are present in about two-thirds of disk galaxies in the local Universe \citep{eskridgeFrequencyBarredSpiral2000,menendez-delmestreNearInfraredStudy2MASS2007,erwinDependenceBarFrequency2018,sahaWhyAreGalaxies2018,leeBarFractionEarly2019}. Observations indicate that the presence and size of these bars correlate to some extent with the stellar mass of the host galaxy. Studies in the near-infrared (NIR) \citep{diaz-garciaCharacterizationGalacticBars2016,erwinDependenceBarFrequency2018} suggest that the fraction of barred galaxies increases with stellar mass, $M_{*}$, particularly in galaxies with $M_{*}<10^{9.7}M_{\odot}$, reaching approximately 50\%-60\% in more massive galaxies. Bar radii vary from less than a kiloparsec (kpc) to 10 kpc in scale. For galaxies with $M_{*}\leq$10$^{10.2}M_{\odot}$, the typical bar size remains relatively constant at around 1.5 kpc. However, in galaxies with higher stellar masses, the size of the bar increases as $M_{*}^{0.56}$ \citep{diaz-garciaCharacterizationGalacticBars2016,erwinDependenceBarFrequency2018,erwinWhatDeterminesSizes2019}.

Bars play an important role in the secular evolution of disk galaxies, especially in their central regions \citep{debattistaBulgesBarsSecular2004,kormendySecularEvolutionFormation2004a,kormendySecularEvolutionDisk2013,cheungGALAXYZOOOBSERVING2013,conseliceEvolutionGalaxyStructure2014a}.
Bars can efficiently redistribute the angular momentum of gas, stars, and dark matter \citep{sellwoodDynamicsBarredGalaxies1993,debattistaDynamicalFrictionDistribution1998,debattistaConstraintsDynamicalFriction2000,athanassoulaMorphologyPhotometryKinematics2002,athanassoulaWhatDeterminesStrength2003a, sellwoodSPIRALINSTABILITIESNBODY2012}.
Bars are able to funnel gas toward the central regions of galaxies, thereby triggering central starbursts \citep{liHYDRODYNAMICALSIMULATIONSNUCLEAR2015,spinosoBardrivenEvolutionQuenching2017,donohoe-keyesRedistributionStarsGas2019,georgeSignificanceBarQuenching2019}.
These may lead to small-scale structures such as nuclear disks and bars \citep{hopkinsHowMassiveBlack2010,Du2015, wozniakHowCanDoublebarred2015,wuMorphologicalKinematicalAnalysis2021}, which can transport gas to yet smaller radii, possibly fueling active galactic nuclei (AGN) \citep{shlosmanBarsBarsMechanism1989, duBlackHoleGrowth2017, liHowNestedBars2023}.
This process may lead to the depletion of gas, causing them to 
quench more quickly \citep{gavazziHa3HaImaging2015,spinosoBardrivenEvolutionQuenching2017,kimStarFormationActivity2017,khoperskovBarQuenchingGasrich2018,linSDSSIVMaNGAIndispensable2020}. 
Moreover, bars buckle spontaneously due to their internal vertical dynamical instabilities forming boxy/peanut-shaped bulges \citep[e.g.,][]{rahaDynamicalInstabilityBars1991,merrittBendingInstabilitiesStellar1994}, which has been observed ongoing in nearby galaxies \citep{erwinCAUGHTACTDIRECT2016,liCarnegieIrvineGalaxySurvey2017}. Therefore, studying bars is crucial in understanding the formation and evolution of galaxies.

Bars form due to internal dynamical instabilities of disks \citep[e.g.,][]{hohlNumericalExperimentsDisk1971,sellwoodDynamicsBarredGalaxies1993,athanassoulaBarHaloInteractionBar2002a,polyachenkoUnifiedTheoryFormation2003,athanassoulaBarFormationEvolution2013a}. Close tidal interactions between galaxies may also trigger, or facilitate, the growth of bars \citep{gerinInfluenceGalaxyInteractions1990,miwaDynamicalPropertiesTidally1998,lokasTIDALLYINDUCEDBARS2016}, although these bars are unlikely be long-lived. Once bars form, they are likely to grow longer and stronger by transferring angular momentum to the dark matter halo \citep{debattistaDynamicalFrictionDistribution1998,debattistaConstraintsDynamicalFriction2000,athanassoulaWhatDeterminesStrength2003a}. Some simulations have shown that bars in gas-rich disk galaxies tend to be weaker and form later than those in gas-poor galaxies, because gas obstructs the growth of the bars by transferring angular momentum to them \citep{berentzenRegenerationStellarBars2004,bournaudLifetimeGalacticBars2005,berentzenGasFeedbackStellar2007,athanassoulaBarFormationEvolution2013a}.

It is still unclear why some disk galaxies are unbarred. \citet{berrierMASSDISTRIBUTIONBAR2016} and \citet{bauerCanStellarDiscs2019} highlighted the difficulty in explaining the existence of unbarred disk galaxies. 
There are two potential explanations for this: either bars in galaxies are destroyed after their formation, or they fail to form at all under certain conditions. Disks are unable to form bars if their dynamical temperature, quantified by the Toomre-$Q$ parameter, is excessively high. $N$-body simulations \citep{athanassoulaBisymmetricInstabilitiesKuz1986, Du2015} proposed a critical value of Toomre-$Q\sim2.2$ for bar formation. \citet{sahaSpinningDarkMatter2013}, however, posited that a spinning dark matter halo could induce bar formation in disks that are otherwise dynamically too hot to form bars on their own. On the other hand, bars are expected to form easily once a dynamically cold disk is assembled. Numerical simulations \citep{shenDestructionBarsCentral2004a,athanassoulaCanBarsBe2005} suggested that an implausibly massive central mass concentration, such as a black hole exceeding $4\%M_*$, would be required to destroy a large-scale bar. Observations find no significant difference in color, rotation, and environment between barred and unbarred galaxies \citep[e.g.,][]{bosmaBarredSpirals1996, berghHOWDIFFERENTARE2011,dengDependenceGalacticBars2023}. Furthermore, if it were the case that the presence of bars in a galaxy is determined by halo dominance, then barred galaxies would have systematically heavier discs than their unbarred cousins. Such a difference would manifest as a systematic offset in the Tully–Fisher relation, since unbarred galaxies of a given luminosity would be predicted to have higher circular speed, which is not observed \citep{mathewsonParameters2447Southern1996,debattistaConstraintsDynamicalFriction2000,courteauTullyFisherRelationBarred2003}. This highlights the mystery of why some galaxies are unbarred.

Bars serve as an important indicator of whether a cosmological simulation can reproduce the dynamical properties of galaxies well. Recent state-of-the-art cosmological hydrodynamical simulations are capable of reproducing reasonable bar structures in galaxies. In the original Illustris cosmological simulation, the fraction of barred galaxies was relatively low (approximately 21\%) among disk galaxies with stellar masses greater than $M_{*}>10^{10.9}M_{\odot}$  \citep{peschkenTidallyInducedBars2019}. The advanced version of Illustris, IllustrisTNG, produces galaxies that successfully emulate real galaxies in many aspects. \citet{zhaoBarredGalaxiesIllustrisTNG2020a} found that the bar fraction in TNG100 is nearly constant at 60\% between $z=0-1$ for $M_{*}>10^{10.6}M_{\odot}$. A similar result was reached in \citet{zhouBarredGalaxiesIllustris12020} and \citet{rosas-guevaraBuildupStronglyBarred2019a} via applying Fourier analysis and selecting somewhat different disk samples. \citet{algorryBarredGalaxiesEAGLE2017} showed that 20\% of the disk galaxies with $M_{*}$=10$^{10.6}$-10$^{11}M_{\odot}$ at $z=0$ in the EAGLE cosmological simulation \citep{schayeEAGLEProjectSimulating2015} have strong bars, while another 20\% have weak bars. In the NewHorizon simulation \citep{duboisIntroducingNEWHORIZONSimulation2021}, \citet{reddishNewHorizonSimulationBar2022} found that there are no massive barred galaxies with $M_{*}>10^{10}M_{\odot}$ at low redshift, attributing this to their large bulges. Therefore, forming well-defined bar structures can also serve as a test for cosmological simulations.

In this study, we use the TNG simulations that offer good statistical samples and realistic representations of bar structures. These enable us to explore the reasons behind why some galaxies have bars while others do not. Moreover, \citet{zhaoBarredGalaxiesIllustrisTNG2020a} and \citet{frankelSimulatedBarsMay2022a} found that there are significantly more short bars in the simulated disk galaxies compared to those observed, requiring further investigation into the prevalence of short bars. 

This paper is organized as follows. In Section~\ref{sec:data}, we introduce the IllustrisTNG simulation and the selection of disk galaxies, as well as the methodology employed in the measurement of bar properties\footnote{The evolution data of bar properties for the selected disk galaxy sample are publicly available at \url{https://www.tng-project.org/lu24}.}. Section~\ref{sec3} defines some parameters and compares barred and unbarred galaxies. In Section~\ref{sec:moremassive}, we explore the evolutionary features of more massive barred galaxies. Section~\ref{sec:lessmassive} discusses the evolutionary characteristics of less massive barred galaxies. Then we focus on the Toomre-$Q$ of barred and unbarred galaxies in Section~\ref{sec:toomre-q}. In Section~\ref{sec:barlength}, we analyze the evolution of bar size in disk galaxies. Finally, we summarize our main conclusions in Section~\ref{sec:conclusion}.

\section{Data and methodology}
\label{sec:data}

\begin{figure*}[t!]
\centering
	\captionsetup[subfigure]{labelformat=empty}
	\subfloat[]{\includegraphics[width=0.49\textwidth]{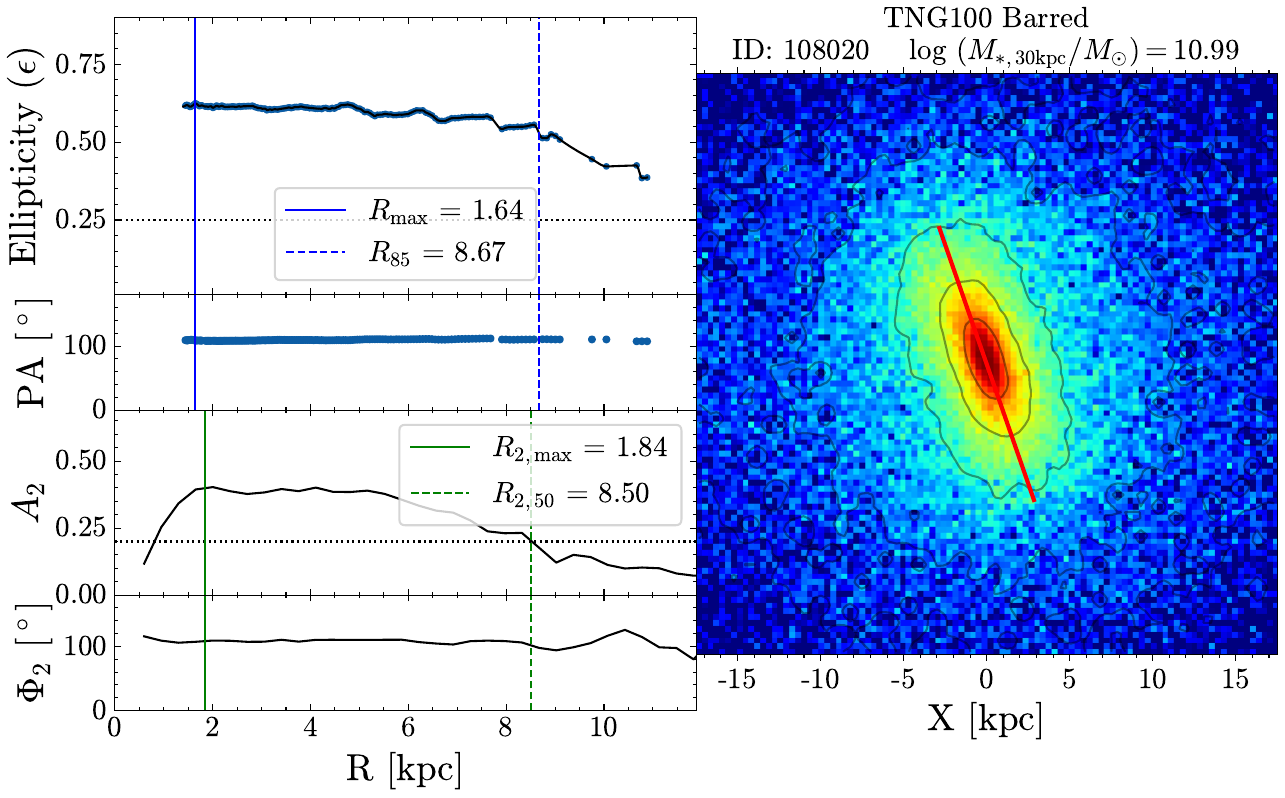}} \hspace{3pt}
 	\subfloat[]{\includegraphics[width=0.49\textwidth]{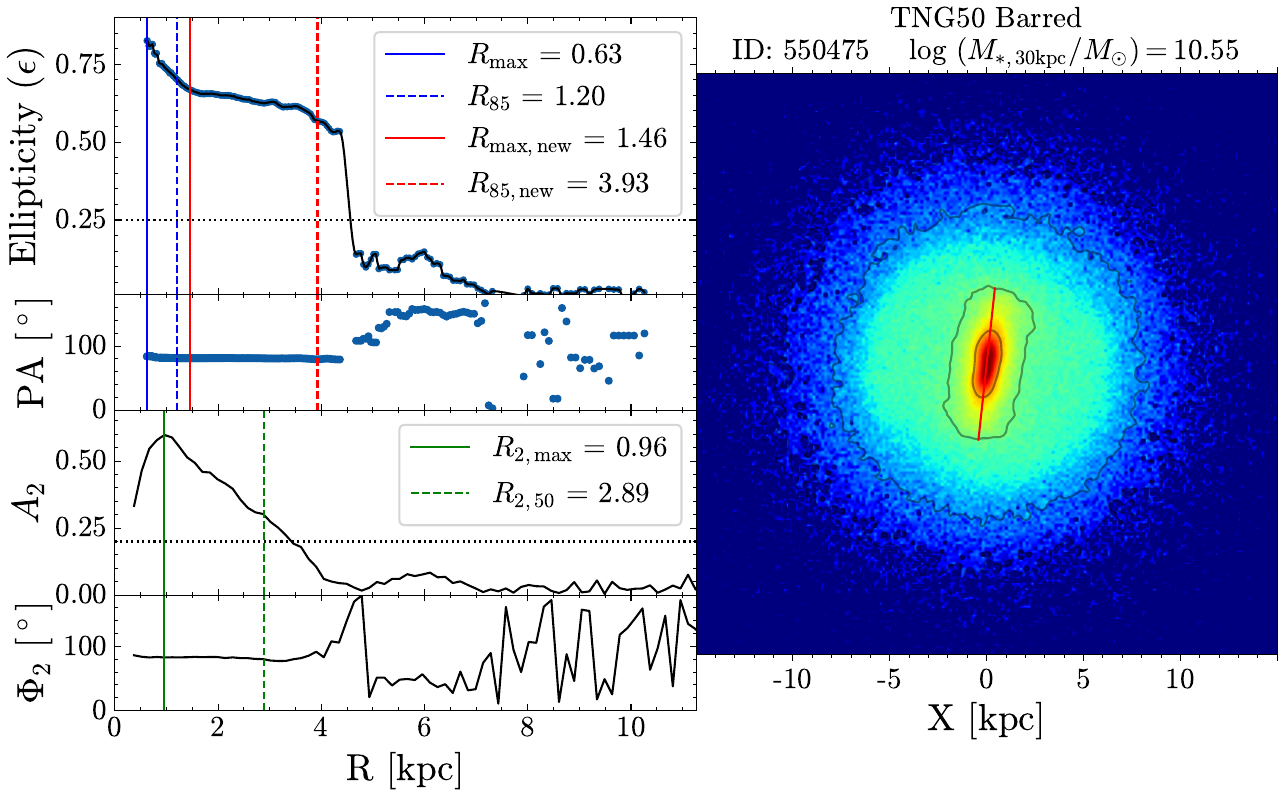}} 
  
  \vspace{-25pt}
    \subfloat[]{\includegraphics[width=0.49\textwidth]{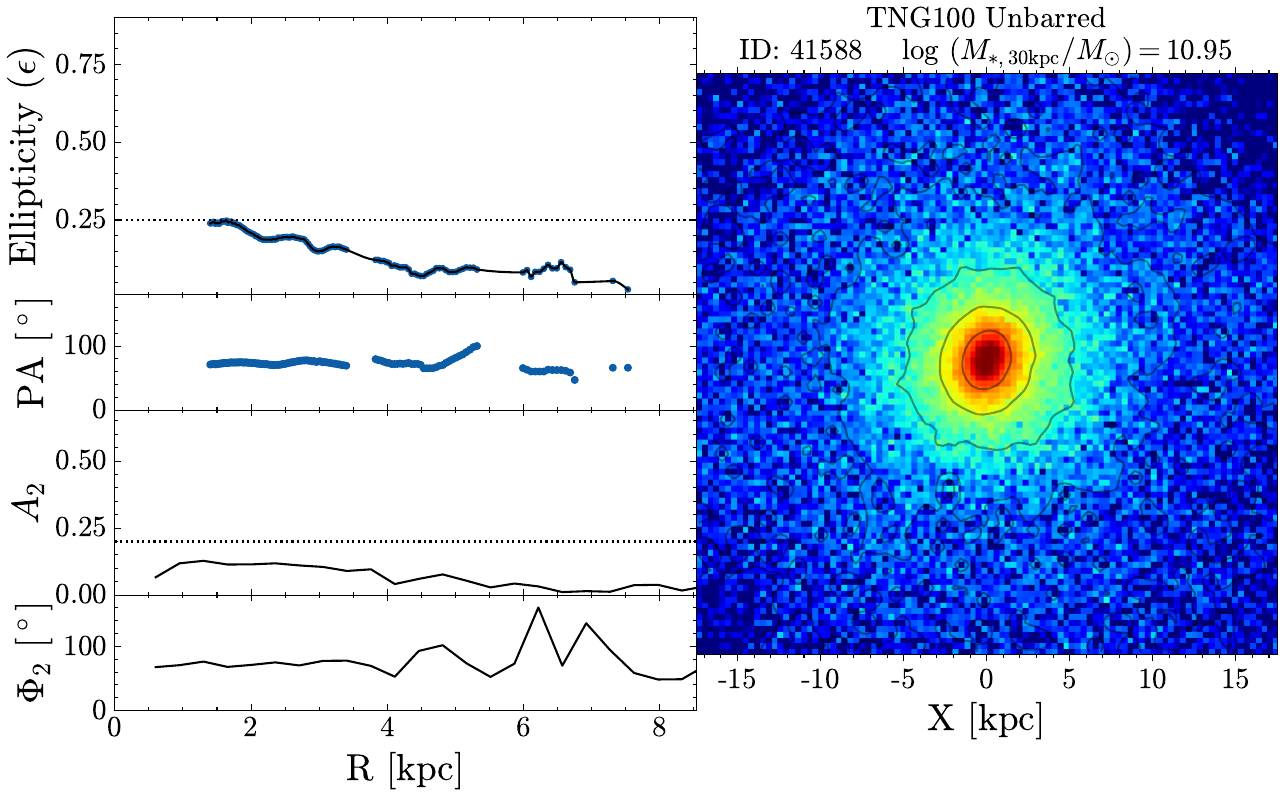}} \hspace{3pt}
    \subfloat[]{\includegraphics[width=0.49\textwidth]{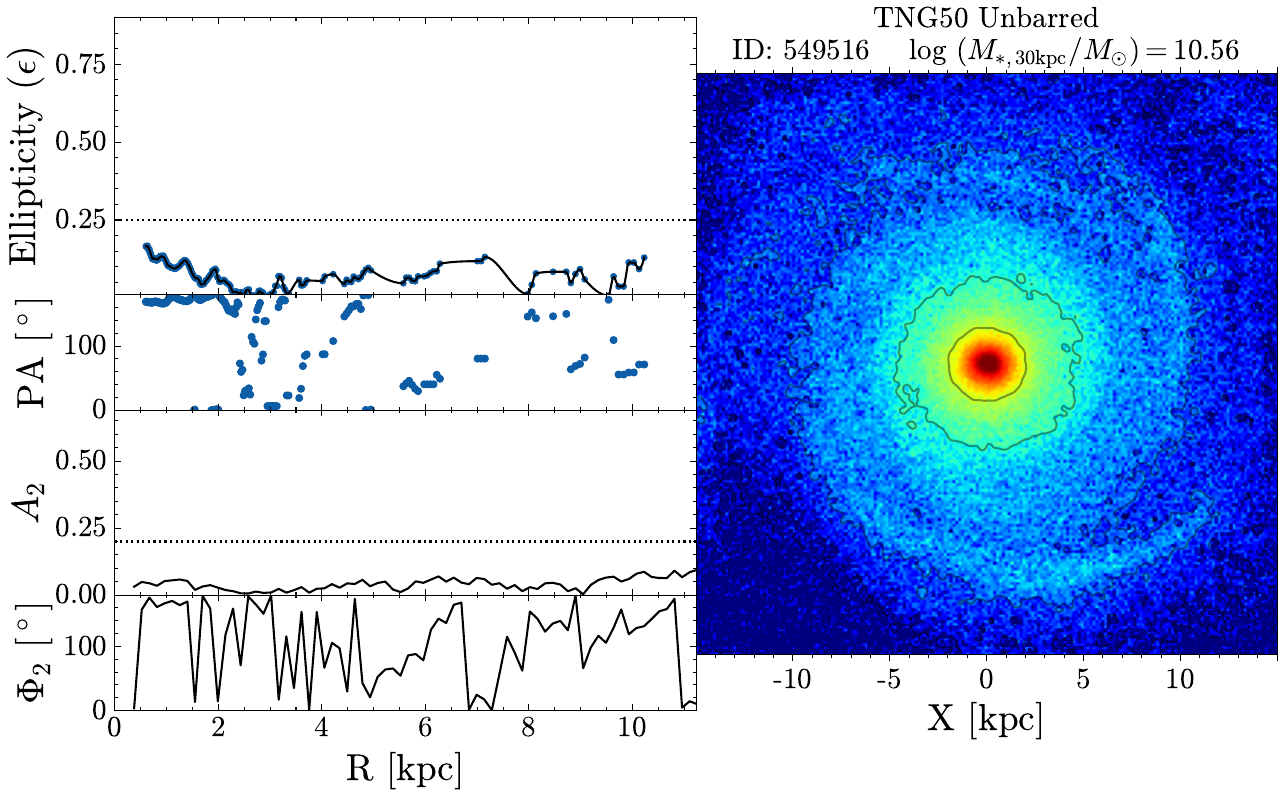}} 
	\caption{Examples of ellipse fitting and Fourier decomposition of barred galaxies (top row) and unbarred galaxies (bottom row) in TNG100 (left) and TNG50 (right). The right-hand column in each panel displays the face-on stellar surface density maps of the galaxies, with each pixel corresponding to $0.35\times0.35\ \mathrm{kpc^2}$ for TNG100, and $0.15\times0.15\ \mathrm{kpc^2}$ for TNG50. The left-hand column of each galaxy presents the radial profiles of ellipticity ($\epsilon$) and position angle (PA) measured via ellipse fitting, along with $A_2$ and $\Phi_2$ from Fourier decomposition. Horizontal dotted lines at $\epsilon =0.25$ and $A_2 =0.2$ are included for reference in each panel. For the barred galaxies: the blue solid and dashed lines represent $R_\mathrm{max}$ and $R_\mathrm{85}$, respectively; for TNG50, the adjusted measurements $R_\mathrm{max,new}$ and $R_\mathrm{85,new}$, calculated with an inner boundary limit of 1.4 kpc, are shown in red; the green solid and dashed line represent $R_\mathrm{2,max}$ and $R_\mathrm{2,50}$, respectively; in the stellar surface density maps, the red line represents the bar size measured by $R_\mathrm{85}$ for TNG100 and $R_\mathrm{85,new}$ for TNG50.}
	\label{fig:example}
 
\end{figure*}

\subsection{IllustrisTNG simulations and disk samples}
\label{sec:illustris}
The IllustrisTNG Project \citep{marinacciFirstResultsIllustrisTNG2018,naimanFirstResultsIllustrisTNG2018,nelsonAbundanceDistributionPhysical2018,nelsonIllustrisTNGSimulationsPublic2019,pillepichFirstResultsIllustrisTNG2018,pillepichFirstResultsTNG502019,springelFirstResultsIllustrisTNG2018} is a set of magnetohydrodynamic simulations of large cosmological volumes run with the moving-mesh code AREPO \citep{springelPurSiMuove2010,pakmorMagnetohydrodynamicsUnstructuredMoving2011,pakmorImprovingConvergenceProperties2016}. The subgrid physics of IllustrisTNG is based on its predecessor, Illustris \citep{vogelsbergerModelCosmologicalSimulations2013,vogelsbergerPropertiesGalaxiesReproduced2014,vogelsbergerIntroducingIllustrisProject2014,genelIntroducingIllustrisProject2014,nelsonIllustrisSimulationPublic2015,sijackiIllustrisSimulationEvolving2015}, but with significant adjustments to the physical models for supernova feedback, active galactic nuclei (AGN) feedback, and chemical enrichment. 

IllustrisTNG has successfully replicated many of the fundamental properties and scaling relationships observed in galaxies. For example, the mass–size relation, as observed in both late-type and early-type galaxies, has been well recovered within observational uncertainties \citep{genelSizeEvolutionStarforming2018,huertas-companyHubbleSequenceIllustrisTNG2019,rodriguez-gomezOpticalMorphologiesGalaxies2019}. Additionally, \citet{rodriguez-gomezOpticalMorphologiesGalaxies2019} made a systematic comparison between the IllustrisTNG simulation and the Pan-STARRS survey, showing that the optical size and shape of the IllustrisTNG galaxies are consistent within $\sim$ 1$\sigma$ scatter of the observed trends. \citet{duOriginRelationStellar2022,duPhysicalOriginMassSize2024} and \citet{maEvolutionaryPathwaysDisk2024} provided a comprehensive explanation of the correlations between stellar mass $M_{*}$, size, and specific angular momentum of stars $j_{*}$ in disk galaxies. Their findings reasonably match with observations of galaxies, offering insights into the physical origins behind these correlations. Furthermore, \citet{xuStudyStellarOrbit2019} found that the fractions of the different orbital components in IllustrisTNG are remarkably consistent with those estimated in CALIFA galaxies \citep{zhuOrbitalDecompositionCALIFA2018}.

We primarily use the TNG100 run, which simulates a volume with a side length of 75 h$^{-1} \approx$ 111 Mpc. For completeness, we also present results from the TNG50 run, which has a side length of 35 h$^{-1} \approx$ 51.7 Mpc. These simulations are publicly available \citep{nelsonIllustrisTNGSimulationsPublic2019}. The TNG100 simulation has an average baryonic resolution element mass of 1.39 $\times$ 10$^{6} M_{\odot}$, whereas the TNG50 simulation has a resolution of 8.5 $\times$ 10$^{4} M_{\odot}$. The gravitational softening length of the stellar particles is 0.5 h$^{-1} \approx$ 0.74 kpc and 0.195 h$^{-1} \approx$ 0.29 kpc in TNG100 and TNG50 for $z<1$, respectively. The dark matter halos and subhalos in each snapshot are derived from the catalog provided by the IllustrisTNG simulation, which identifies bound substructures using a Friends-of-Friends (FoF) group finding algorithm \citep{davisEvolutionLargescaleStructure1985} and then a SUBFIND algorithm \citep{springelPopulatingClusterGalaxies2001,dolagSubstructuresHydrodynamicalCluster2009} and tracked over time by the SUBLINK merger tree algorithm \citep{rodriguez-gomezMergerRateGalaxies2015}. A galaxy is defined as a gravitationally bound object with at least 100 stellar particles within a given halo or subhalo. The central galaxy (subhalo) is the first (most massive) subhalo of each FoF group, while the other galaxies within the FoF halo are its satellites.

The TNG100 disk galaxies used in this study are from the disk galaxies catalog \footnote{\url{https://www.tng-project.org/files/TNG_BarProperties/tng100-1_099_bars_zhao20.hdf5}} selected by \citet{zhaoBarredGalaxiesIllustrisTNG2020a}, who selected 3866 disk galaxies in the TNG100 simulation with a stellar mass within 30 kpc $M_\mathrm{*,30kpc}\ge 10^{10}\ M_{\odot}$, and then identified 1182 barred galaxies using ellipse fitting. These disk galaxies were selected based on the kinematic parameter $\kappa_\mathrm{rot,30kpc} \geqslant$ 0.5, where $\kappa_\mathrm{rot,30kpc}$ is the $\kappa_\mathrm{rot}$ calculated within 30 kpc. The definition of $\kappa_\mathrm{rot}$ can be found in Table~\ref{tab:parameterdefine}. Galaxies selected by this criterion clearly exhibit a disk morphology. To ensure a sufficient sample of disk galaxies and barred galaxies, we select disk galaxies in the stellar mass range of log$(M_\mathrm{,30kpc}/M_{\odot}) = 10.4 - 11.1$ from TNG100. Additionally, following the same criterion, we also select 615 disk galaxies with log$(M_\mathrm{*,30kpc}/M_{\odot}) = 10 - 11.1$ from TNG50. These galaxies form our parent sample of disk galaxies; The evolution of their bar sizes is investigated in Section~\ref{sec:barlength}. All galaxies in our sample are centered at the minimum position of their potential well and then are rotated to the face-on view based on the stellar angular momentum of all stars within a spherical radius of $r<8$ kpc to accurately measure their properties.

\subsection{Measurement of bars using both ellipse fitting and Fourier analysis}
\label{sec:barcharacter}

\begin{figure*}
 
\centering
	\captionsetup[subfigure]{labelformat=empty}
	\subfloat[]{\includegraphics[width=0.828\columnwidth]{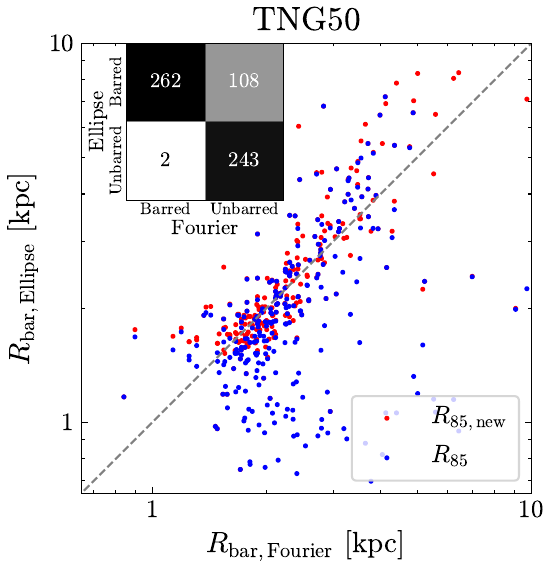}}
 	\subfloat[]{\includegraphics[width=0.81\columnwidth]{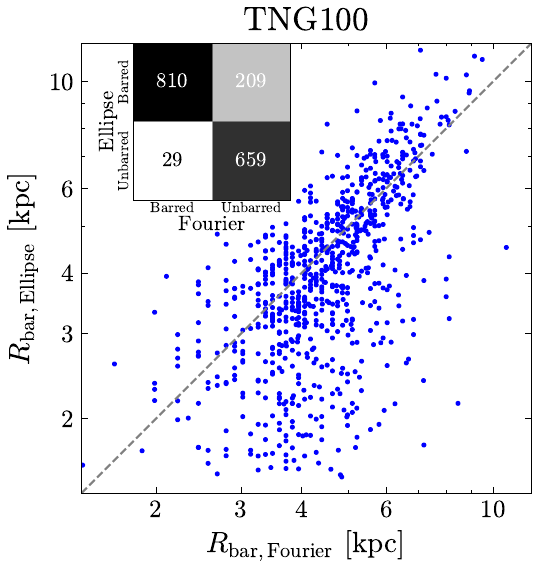}} 
    \vspace{-10pt}
	\caption{Comparison of bar sizes measured by Fourier decomposition (x-axis) and ellipse fitting (y-axis) for TNG50 (left panel) and TNG100 (right panel). For TNG50, bar sizes obtained from ellipse fitting include the original $R_{85}$ (blue points) and the adjusted $R_{85,\mathrm{new}}$ (red points). The gray dashed lines represent the 1:1 reference lines in both panels. The confusion matrix in the top left corner of each panel compares the number of bars identified by Fourier decomposition and ellipse fitting. Points in the plot represent galaxies identified as barred by both methods.}
	\label{fig:fouriervsellipse}

\end{figure*}

\subsubsection{Fourier decomposition}
The $m=2$ term of the Fourier expansion $A_2$ is commonly used to quantify bar structures in simulations \citep[e.g.,][]{athanassoulaMorphologyPhotometryKinematics2002, Du2015, rosas-guevaraBuildupStronglyBarred2019a}. $A_2$ and its phase $\Phi_2$ are defined as:
\begin{equation}
    A_{2}(R)=\frac{|\sum_{j}m_{j}e^{2i\theta_j}|}{\sum_{j}m_j}
\end{equation}
\begin{equation}
    \Phi_2(R) = \frac{1}{2}\arctan[\frac{\sum_{j}m_{j}\sin(2\theta_j)}{\sum_{j}m_{j}\cos(2\theta_j)}]
\end{equation}

These definitions apply to face-on galaxy projections, where $m_j$ and $\theta_j$ are the mass and azimuthal angle of each star particle. The summations are performed over all stellar particles within the annular region of width ${\rm d}R$ at a radius $R$ from the galactic center and being coaxial with it. The bin width is chosen to be half the gravitational softening length. Numerous criteria and methods have been devised for identifying and measuring bars \citep[e.g.,][]{athanassoulaMorphologyPhotometryKinematics2002,erwinHowLargeAre2005,zhouBarredGalaxiesIllustris12020,rosas-guevaraEvolutionBarredGalaxy2022a,Anderson2024}. We adopt criteria and methods similar to those in  \citet{Anderson2024} to characterize bars: (1) the maximum value of $A_2$ within the bar, $A_\mathrm{2,max}$, must exceed 0.2, and the phase $\Phi_2$ should vary by less than $10^{\circ}$ (2) The bar's radius, $R_\mathrm{2,50}$, is defined as the cylindrical radius where $A_2$ drops to half of $A_\mathrm{2,max}$. Additionally, we define the bar strength as $A_\mathrm{2,max}$, and $R_\mathrm{2,max}$ is the radius of $A_\mathrm{2,max}$. Figure ~\ref{fig:example} presents two examples of barred (top) and unbarred (bottom) galaxies from TNG100 (left) and TNG50 (right). The corresponding radial profiles of $A_2$ and $\Phi_2$ are displayed.   

\subsubsection{Ellipse fitting}
\label{ellipsefitting}
Ellipse fitting has been widely used in measuring bar in observations. We perform ellipse fitting to identify bars, as suggested by \citet{zhaoBarredGalaxiesIllustrisTNG2020a}. Utilizing the \texttt{photutils} package \citep{larry_bradley_2023_7946442} in Python, we employ the standard iterative ellipse fitting method of \citet{jedrzejewskiCCDSurfacePhotometry1987}. Stellar particles are binned into square bins of $\mathrm{0.35\times0.35\ kpc^2}$ for TNG100 and $\mathrm{0.15\times0.15\ kpc^2}$ for TNG50. Notably, the length of the bin side is roughly equal to half of the gravitational softening radius. We fit ellipses to the isodensity contours of the face-on surface density maps of disk galaxies in TNG100 and TNG50. For each fit, this method measures the radial profiles of ellipticity $\epsilon$, position angle (PA), and semi-major axis length (SMA). We set the minimum SMA of ellipse fitting to a length of 4 pixels, corresponding to 1.4 and 0.6 kpc for TNG100 and TNG50, respectively, as suggested by \citet{zhaoBarredGalaxiesIllustrisTNG2020a} to reduce the large uncertainty of the ellipse fitting in the central regions.

We apply the same criteria for identifying bars as suggested by \citet{martinez-valpuestaEvolutionStellarBars2006a} and \citet{zhaoBarredGalaxiesIllustrisTNG2020a}: (1) the maximum value of $\epsilon$ ($\epsilon_\mathrm{max}$) within the bar must be greater than 0.25, and the PA should vary by less than $10^\circ$; (2) $\epsilon$ should decrease by more than 0.1 from the maximum value outward. Following these criteria, we define bar strength as $\epsilon_\mathrm{max}$, $R_\mathrm{max}$ as the radius at $\epsilon_\mathrm{max}$, and $R_\mathrm{PA}$ as the maximum radius where the PA variation remains $<10^\circ$.

\citet{zhaoBarredGalaxiesIllustrisTNG2020a} found that $R_\mathrm{85}$, the radius where $\epsilon$ declines to 85\% of $\epsilon_\mathrm{max}$, generally aligns well with visual estimates of bar sizes in TNG100 galaxies. However, in our analysis of TNG50 galaxies, $R_\mathrm{85}$ has a tendency to underestimate bar sizes due to the significant increase in $\epsilon$ in the central regions of galaxies, which is an unusual phenomenon in observed galaxies as well as TNG100. It is not clear what causes such a sharp enhancement of bar ellipticity in TNG50. The top-right panel of Figure \(\ref{fig:example}\) shows this effect in a TNG50 barred galaxy. Such inner higher ellipticity can be suppressed by the effects of the point-spread function \citep{goncalvesEllipsefittingMockImages2024}. However, we adopt a simpler approach here. Specifically, this problem can generally be avoided by measuring $\epsilon_\mathrm{max}$ at $R > 1.4$ kpc where $\epsilon$ is nearly constant over a certain region. We verified that this 1.4 kpc cutoff effectively excludes most of the central regions exhibiting a sharp increase in ellipticity, allowing us to reliably identify the location of the ellipticity drop at the bar's outer edge. Consequently, we assess bars in TNG50 using ellipse fitting data for radii greater than 1.4 kpc in the case of a normal long bar. In the case of short bars, $\epsilon_\mathrm{max}$ is measured by taking all radii into account. The adjusted $R_\mathrm{85,new}$ offers a more accurate measurement of bar size in especially massive galaxies with long bars.

We perform ellipse fitting and Fourier decomposition on the stellar surface density of TNG50 disk galaxies with $M_\mathrm{*,30kpc}=10^{10.0}-10^{11.1}\ M_{\odot}$ and TNG100 disk galaxies with $M_\mathrm{*,30kpc}=10^{10.4}-10^{11.1}\ M_{\odot}$. Figure~\ref{fig:fouriervsellipse} compares the results obtained from these two methods. Overall, the bar sizes derived from ellipse fitting and Fourier decomposition are well-correlated. For TNG50, the original $R_{85}$ measurements (blue points) show many values clustering around $R_{\rm bar,Ellipse}=1$ kpc, underestimating bar sizes. In contrast, the corrected $R_{85,\mathrm{new}}$ aligns better with the results from Fourier decomposition.
As shown in the confusion matrix, ellipse fitting identifies more bars than Fourier decomposition, consistent with earlier studies suggesting that ellipse fitting is better at detecting weak bars \citep{leeBarFractionEarly2019,zhaoBarredGalaxiesIllustrisTNG2020a}. It is worth mentioning that the strength of bars in TNG100 is likely to be weakened partly due to numerical issues. This results in a smaller \(A_{2,\mathrm{max}}\), which in turn causes \(R_{2,\mathrm{max}}\) to have relatively larger values in weak bar cases than their measurements using ellipse fitting. Additionally, the variation of $\epsilon$ is smaller towards around the ends of bars, as shown in Figure~\ref{fig:example}, though the trends of both $A_2$ and $\epsilon$ with radius are similar. Therefore, the ellipse fitting method used in this study is likely to provide a more reliable bar detection. Hereafter, we refer to bar size as $R_\mathrm{bar}$, which corresponds to $R_{85}$ for TNG100 and the corrected $R_{85,\mathrm{new}}$ for TNG50. Note that $R_\mathrm{bar}$ does not exceed $R_\mathrm{PA}$.

\subsubsection{Identification and classification of bars}
\label{sec:Sample} 

\begin{figure}
\centering
	\includegraphics[width=0.9\columnwidth]{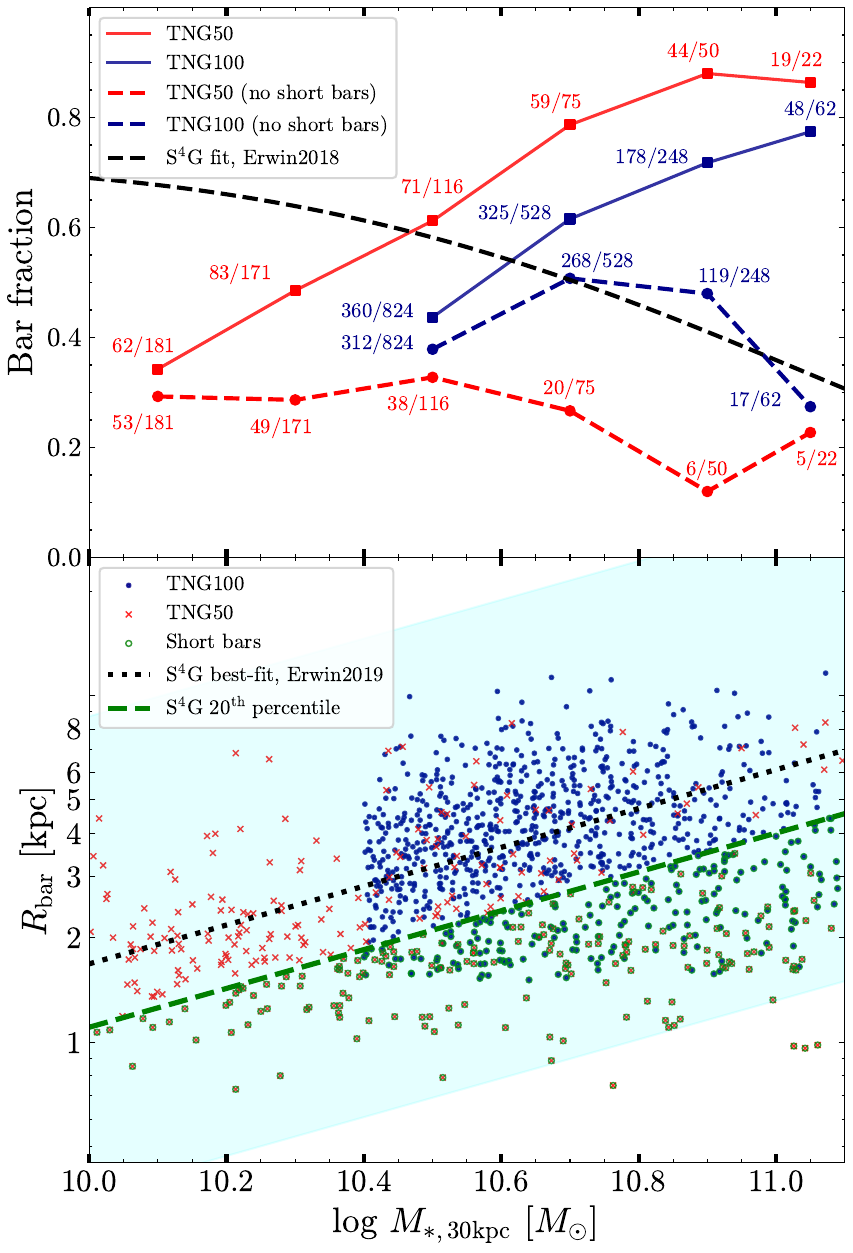}
	\caption{Variations of bar fraction, $f_\mathrm{bar}$, (upper) and distributions of bar size, $R_\mathrm{bar}$, (lower) with stellar mass $M_\mathrm{*,30kpc}$ for galaxies in TNG100 (blue) and TNG50 (red). In the bottom panel, the cyan shaded region represents the 3$\sigma$ scatter range of bar sizes in the S$^{4}$G survey \citep{erwinDependenceBarFrequency2018}. The black dotted line shows the best-fit relationship between bar size and stellar mass for the S$^{4}$G sample \citep{erwinWhatDeterminesSizes2019}. The green dashed lines indicates the 20th percentile of the bar size distribution derived from the S$^{4}$G sample. Galaxies with short bars (bar sizes below the S$^{4}$G 20th percentile) in TNG50 and TNG100 are highlighted with green circles. In the top panel, the solid and dashed lines show the bar fraction for all bars and bars longer than the S$^{4}$G 20th percentile for TNG50 (red) and TNG100 (blue), respectively. The numbers next to the points indicate the number of barred galaxies over the total number of disk galaxies in each stellar mass bin for TNG50 and TNG100. Black line represent the fit of bar fractions observed in the near-infrared S$^{4}$G survey \citep{erwinDependenceBarFrequency2018}.}
	\label{fig:TNG50vsTNG100vsS4Gselect}
\end{figure}

We use the bar size measurements obtained through ellipse fitting. Furthermore, we visually inspect galaxies with weak bars and find that approximately 10\% of the barred galaxies are misclassified as having weak bars due to irregular morphology in their central regions. After removing these cases, our final sample includes 338 barred galaxies in TNG50 and 911 barred galaxies in TNG100.

The bottom panel of Figure~\ref{fig:TNG50vsTNG100vsS4Gselect} shows the distribution of bar sizes, $R_\mathrm{bar}$, and galaxy mass, $M_\mathrm{*,30kpc}$, for both TNG50 and TNG100. Nearly all the data points fall within the 3$\sigma$ region (cyan area) of the S$^{4}$G sample \citep{erwinWhatDeterminesSizes2019}. Observations find a strong relationship between bar size and galaxy mass $R_\mathrm{bar} \propto M_{*}^{0.56}$ (the black dotted line) for galaxies with stellar masses greater than $10^{10.1} M_{\odot}$ \citep{erwinWhatDeterminesSizes2019}. However, in the TNG samples, at the high-mass end ($M_{*}>10^{10.6} M_{\odot}$), the bar sizes do not increase much with mass; in fact, there is even a slight decrease in bar sizes. This can be accounted for by the potential overproduction of short bars in TNG simulations. It also results in excess of the bar fraction in the high-mass region of TNG100 compared to observations (the blue dotted line in the top panel of Figure~\ref{fig:TNG50vsTNG100vsS4Gselect}). Additionally, a possible observational bias exists, where short bars may be missed when they coexist with a massive bulge. We use the 20th percentile line from S$^4$G to classify short bars (dashed green line in Figure~\ref{fig:TNG50vsTNG100vsS4Gselect}), which selects the same short bar sample as in \citet{zhaoBarredGalaxiesIllustrisTNG2020a}. After excluding the short-bar sample, the bar fraction at the high-mass end of TNG100 aligns with the bar fraction from S$^4$G (the blue-dotted line). In TNG50, the excess of the bar fraction at the high-mass end is more evident. \citet{roshanFastGalaxyBars2021} also discovered that the bars in TNG50 are approximately 40\% shorter than those in TNG100. Moreover, \citet{frankelSimulatedBarsMay2022a} demonstrated that, on average, the bars in TNG50 are around 35\% shorter than those observed in MaNGA, which is consistent with our findings.

Although the IllustrisTNG simulations do not perfectly reproduce the observations, as this is a challenging task, excluding the short bars results in the bar fraction in TNG100 aligning well with the observations from S$^4$G. This provides a valuable framework for understanding the bar fraction trend and the factors that influence it. In this work, we select 838 disk galaxies with $M_\mathrm{*,30kpc}=10^{10.6}-10^{11.1}\ M_{\odot}$ at $z=0$ in TNG100. We then categorize these disk galaxies into three groups:
\begin{description}  
\item [$\bullet$ barred galaxies:] 404 disk galaxies with normal bars,
\item [$\bullet$ short-bar galaxies:] 147 disk galaxies with bars shorter than the 20th percentile line from S$^4$G observation,
\item[$\bullet$ unbarred galaxies:] 287 disk galaxies without any bar.
\end{description}
We primarily investigate the evolutionary histories of these galaxies and the differences among the three groups. Galaxies in TNG50 (red crosses in Figure~\ref{fig:TNG50vsTNG100vsS4Gselect}) are included to extend to lower masses and compare them with TNG100's results. Finally, in Section~\ref{sec:barlength}, we study the evolution of bar sizes for the full disk galaxy sample, which includes disk galaxies with $M_\mathrm{*,30kpc}=10^{10.4}-10^{11.1}\ M_{\odot}$ from TNG100 and those with $M_\mathrm{*,30kpc}=10^{10.0}-10^{11.1}\ M_{\odot}$ from TNG50.

\section{Correlation analysis between the properties of galaxies and bars}
\label{sec3}

\subsection{Definition of parameters}
\label{sec:parameter}

\begin{table}
\caption{Parameter definitions.} \label{tab:parameterdefine}

\begin{tabular}{|m{6em}<{\centering}|m{16em}|}
\hline
 Symbol [Unit] & Definition   \\
 \hline
$R_e$ & The radius containing half of the stellar mass $M_\mathrm{*,30kpc}$ in the cylindrical coordinate.\\
\hline
$h_{R}$ & The disk scale length, derived as described in Apppendix \ref{sec:parameter h}.\\
  \hline
 $C_{82}$ & The ratio of the cylindrical radius containing 80\% of the stellar mass, $R_\mathrm{80}$, to the cylindrical radius containing 20\% of the stellar mass, $R_\mathrm{20}$. \\
\hline
 $\Sigma_\mathrm{1kpc}$, $\Sigma_{R_{e}}$ [$M_{\odot}/{\rm kpc}^2$]& The surface stellar mass density measured within 1 kpc ($R_{e}$).   \\
 \hline
 $\kappa_\mathrm{rot}$ & The stellar kinetic energy fraction in ordered rotation of all stellar particles \citep{salesFeedbackStructureSimulated2010}, $\sum mv_\mathrm{\phi}^{2} / \sum mv^{2}$, where $m$, $v_\mathrm{\phi}$, and $v$ are mass, azimuthal velocity, and total velocity of each stellar particle.  \\
\hline
 $f_\mathrm{gas}$ & The mass ratio between cold gas (HI+H$_{2}$) and $M_*$. \tablefootmark{a} \\
 \hline
$f_\mathrm{BH}$ & The mass ratio between the black hole and $M_*$.  \\
\hline
$f_\mathrm{baryon}$ & The baryonic mass ratio, $M_\mathrm{baryon}/M_\mathrm{tot}$.   \\
\hline
 Z$_{*}$ & The stellar metallicity. \\
\hline
 SFR $[M_{\odot}/yr]$, (sSFR $[{\rm Gyr}^{-1}]$) & The (specific) star formation rate.  \\
 \hline
$f_\mathrm{ex\ situ}$ & The ex-situ mass ratio. \tablefootmark{b}\\
\hline
$f_\mathrm{bulge}$, $f_\mathrm{disk}$, $f_\mathrm{halo}$ & The mass ratios of kinematically-derived bulges, disks, and stellar halos in galaxies. $f_\mathrm{halo}$+ $f_\mathrm{disk}$+ $f_\mathrm{bulge}=1$ in each galaxy. \tablefootmark{c}  \\
\hline
$\lambda_*$ & The dimensionless stellar spin parameter, $j_{*}\ /\ M_{*}^{0.55}$, defined based on the $j_{*}-M_{*}$ relation of disk galaxies from TNG simulations measured by \citet{duOriginRelationStellar2022,duPhysicalOriginMassSize2024} in order to remove the dependence on $M_{*}$. \\
\hline
$c_{200}$ & The dark matter halo concentration, $c_{200} = r_{200}/r_{-2}$, where $r_{200}$ is the viral radius and $r_{-2}$ is the scale radius, derived from fitting the halo density profile by the \citet{einastoConstructionCompositeModel1965} profile.\\
\hline
\end{tabular}

\tablefoot{
These parameters are measured for all particles of a SUBFIND galaxy in the face-on view within a cylindrical coordinate system. No radial limit is imposed, with the exception of \(\kappa_{\mathrm{rot},30\mathrm{kpc}}\).}
\tablefoottext{a}{Cold gas mass from \citep{diemerModelingAtomictomolecularTransition2018,diemerAtomicMolecularGas2019}.
} \tablefoottext{b}{The ex-situ mass from \citet{rodriguez-gomezStellarMassAssembly2016}.} 
 \tablefoottext{c}{kinematically-derived components from \citet{duIdentifyingKinematicStructures2019,duKinematicDecompositionIllustrisTNG2020}.
}
\end{table}

Table~\ref{tab:parameterdefine} lists the parameters that we have examined aiming to investigate the difference between barred, unbarred, and short-bar galaxies. 
$\Sigma_\mathrm{1kpc}$ quantifies the central mass density of galaxies within 1 kpc. It has been considered as a good indicator of bulges and black hole growth \citep[e.g.,][]{fangLINKSTARFORMATION2013,niDoesBlackHole2019,niRevealingRelationBlack2021}, although TNG simulations cannot resolve the central regions of galaxies well.
Previous studies have shown that mass concentration affects the presence of bars. Parameters such as $\Sigma_{R_\mathrm{e}}$ and $C_\mathrm{82}$ quantify the overall concentration of galaxies, and $c_{200}$ measures the concentration of dark matter halos. Barred and unbarred galaxies differ in terms of their galaxy size and disk size \citep{sanchez-janssenEvidenceSecularEvolution2013,erwinWhatDeterminesSizes2019}. We use $R_e$ to measure the size of galaxies and $h_R$ to represent the size of galaxy disks. $\kappa_\mathrm{rot}$ quantifies the relative strength of cylindrical rotation. In addition, $f_\mathrm{ex\ situ}$, characterizes the strength of mergers of galaxies during their evolution and is adopted from \citet{rodriguez-gomezStellarMassAssembly2016}. \citet{rodriguez-gomezStellarMassAssembly2016} suggested that on average, $\sim$50\% of the ex-situ stellar mass comes from major mergers (stellar mass ratio $\mu>1/4$), $\sim$20\% from minor mergers ($1/10 <\mu<1/4$), $\sim$20\% from very minor mergers ($\mu<1/10$), and the remaining $\sim$10\% from stars that were stripped from surviving galaxies (e.g., flybys or ongoing mergers). We adopt the mass ratio of kinematically-derived structures in \citet{duKinematicDecompositionIllustrisTNG2020} and \citet{duIdentifyingKinematicStructures2019}\footnote{The kinematic decomposition data are publicly available at \url{https://www.tng-project.org/data/docs/specifications/#sec5m}.} where an automated Gaussian Mixture Model (auto-GMM) is used to decompose simulated galaxies in a phase space of circularity, binding energy, and non-azimuthal angular momentum \citep[see also][]{Abadi2003b, Domenech-Moral2012, Obreja2016, Obreja2018a}. The disk structures are composed of stellar particles with strong or moderate rotation derived using the kinematic method. This method is able to break the morphological degeneracy between bulges and stellar halos \citep{duKinematicDecompositionIllustrisTNG2020}. \citet{Du_2021} suggested that only $f_\mathrm{halo}$ is tightly correlated with external processes, i.e., nurture, which is similar to $f_\mathrm{ex\ situ}$. The mass ratio of kinematically-defined bulges $f_{\rm bulge}$ is closely associated with the early gas-rich assembly of galaxies in the early Universe.

\subsection{Difference between barred and unbarred galaxies}
\label{sec:KS}
\begin{figure*}

    \centering 	                   \includegraphics[width=0.8\textwidth]{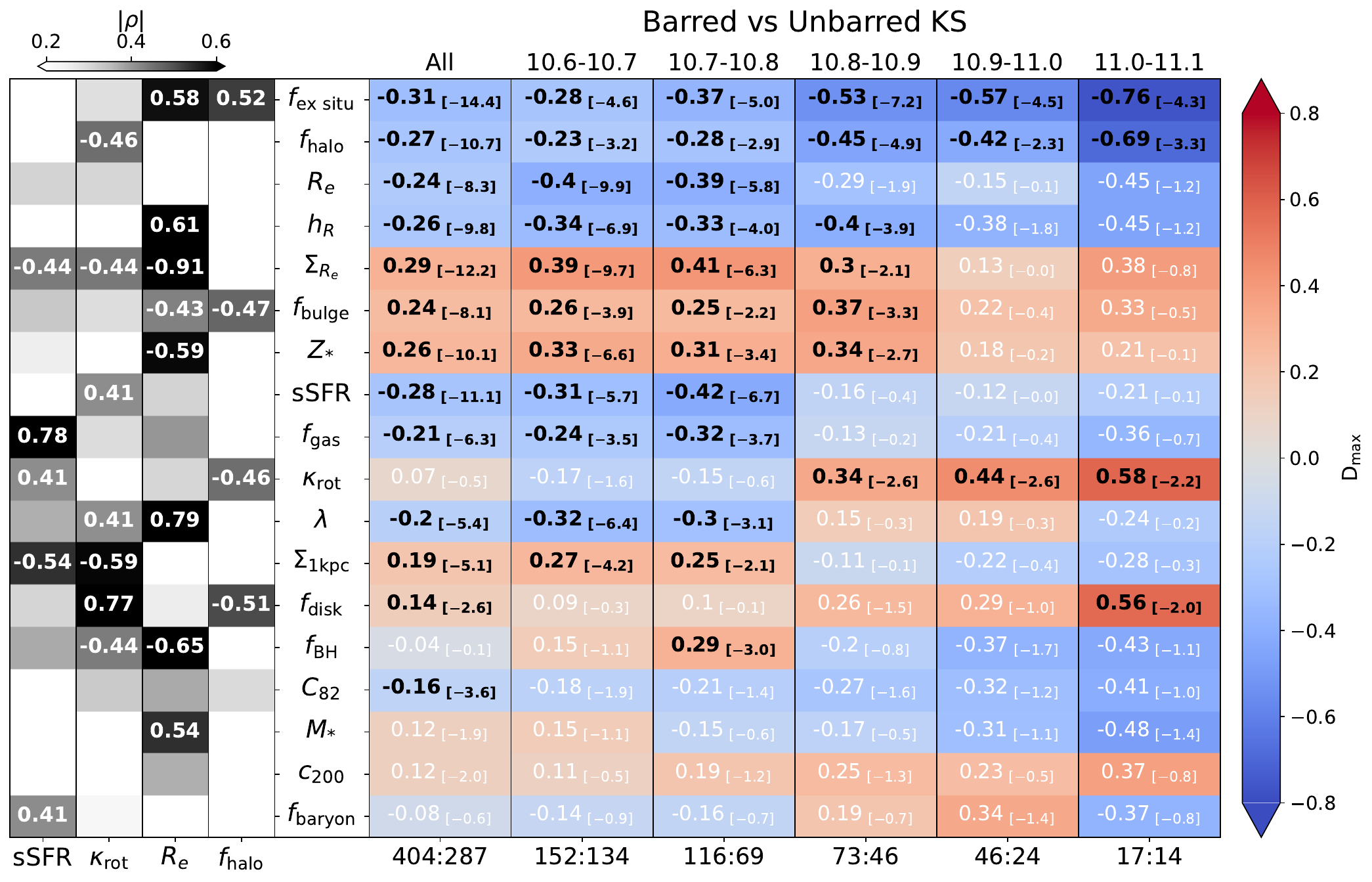} 
        \caption{Results of the KS test (right) and Pearson correlation coefficient (left) between various parameters of barred and unbarred galaxies. In the right part of the graph (in red or blue), the leftmost column represents the KS test for all samples, while the remaining five columns represent the KS-test results for samples in five different mass ranges. The two numbers at the bottom of each column give the sample sizes involved in the KS test, with the number on the left corresponding to the count of barred galaxies, and the number on the right to that of unbarred galaxies. Within each cell, the larger number on the left represents $D_\mathrm{max}$, the maximum distance between the two cumulative distribution functions of the two samples. A positive $D_\mathrm{max}$ indicates barred galaxies have systematically larger corresponding parameter values, whereas a negative $D_\mathrm{max}$ indicates the opposite. The smaller number in the bracket gives the log p-value. These numbers are highlighted in bold black when the p-values are less than 0.01, indicating significant differences. The ordering of parameters is obtained by the descending order of the sum of $|D_\mathrm{max}|$ in all mass bins. In the left part of the figure (in blue), four columns show the Pearson correlation coefficients ($\rho$) between the four parameters (sSFR, $\kappa_\mathrm{rot}$, $R_e$, and $f_\mathrm{halo}$) and all the parameters, the values on each cell being the corresponding $\rho$, labeled only for $|\rho|>0.4$. The color on each cell is determined by $|\rho|$, with a darker color indicating a higher correlation between the two parameters. In the analysis here, these parameters are performed in the log space: $R_e$, $\Sigma_{R_e}$, sSFR, SFR, $f_\mathrm{gas}$, $\lambda_*$, $\Sigma_\mathrm{1kpc}$, $f_\mathrm{BH}$, $M_*$, and $f_\mathrm{baryon}$.}\label{fig:barunKS_score}
\end{figure*}

\begin{figure*}[t]
	\centering 	                   \includegraphics[width=0.9\textwidth]{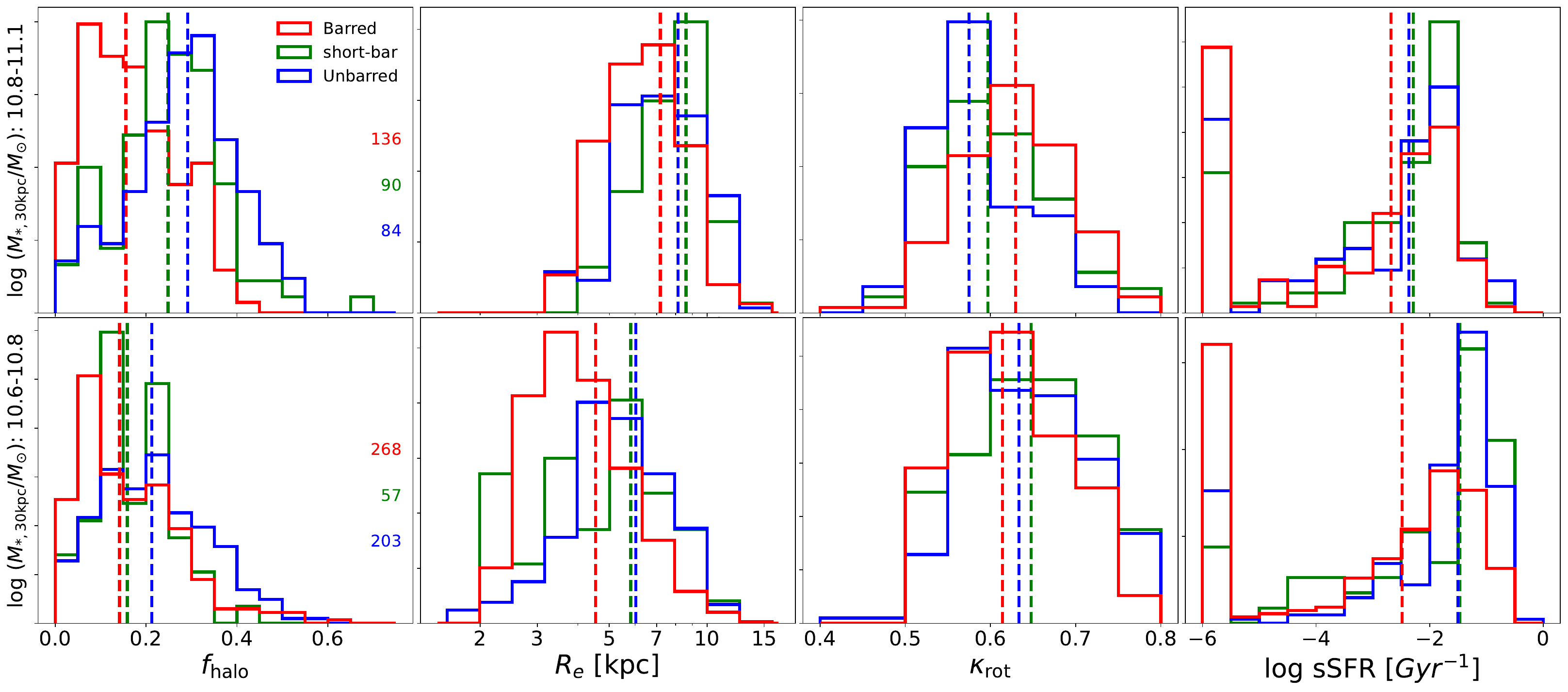}
	\caption{Distributions of $f_\mathrm{halo}$, R$_{e}$, $\kappa_\mathrm{rot}$, and sSFR in barred (red), short-bar (green), and unbarred (blue) galaxies in TNG100. The lower and upper panels show relatively less massive and more massive galaxies, respectively. The dashed lines of the corresponding color indicate the median values. The numbers shown in the left column give the number of galaxies for the corresponding galaxy types and mass ranges. }
	   \label{fig:sixdistribu}
\end{figure*}

We use the two-sample Kolmogorov-Smirnov (KS) test to quantify the difference of parameters listed in Table \ref{tab:parameterdefine} between barred and unbarred galaxies, as shown in the right part of Figure~\ref{fig:barunKS_score}. $D_\mathrm{max}$ measures the maximum difference of the cumulative distribution between the two samples, while the p-value at the right-bottom corner gives the probability of the test statistic under the null hypothesis that the two distributions are identical. Thus, a smaller p-value indicates a higher probability for two data sets to come from different distributions.
The KS test is performed on a sample that includes 417 barred galaxies and 307 unbarred galaxies at redshift  $z=0$ in TNG100. Galaxies with short bars are not included here in order to simplify our analysis. The sample is divided into five mass bins to minimize any potential effect from $M_*$. Whenever a significant difference is found (p-value < 0.01), we highlight \( D_\mathrm{max} \) values and p-values in bold black. In some cases, it is not immediately clear whether the two parameters are independent. Therefore, we use four parameters $f_\mathrm{halo}$, $R_e$, 
$\kappa_\mathrm{rot}$, and sSFR to represent the fundamental properties of the galaxy. They quantify the strength of mergers, size, rotation, and star formation, respectively. We calculate the Pearson correlation coefficients ($\rho$) of these four parameters with the others (the left part of Figure~\ref{fig:barunKS_score}). A larger $|\rho|$ represents a stronger linear correlation. Almost all of the parameters have moderate or strong correlations with at least one of the four parameters. 

Figure~\ref{fig:barunKS_score} shows that the influence of nurture, specifically mergers and close tidal interactions between galaxies, plays a significant role in determining the presence of bars, especially in massive disk galaxies. There are large differences in both $f_\mathrm{ex\ situ}$ and $f_\mathrm{halo}$ between barred and unbarred galaxies, indicated by the large $D_\mathrm{max}$ across all mass bins. This disparity is more pronounced in more massive cases. 

The influence of natural factors becomes as important as nurture in the relatively less massive disk galaxies with log \((M_\mathrm{*,30kpc}/M_\odot) = 10.6-10.8\). Studies by \citet{duPhysicalOriginMassSize2024} and \citet{maEvolutionaryPathwaysDisk2024} have established empirical correlations in galaxies where mergers have only slightly affected their evolution, i.e., galaxies dominated by nature or internal processes. Their results suggest that strong correlations among $M_*$, angular momentum (quantified by \(\lambda_*\) here), \(R_{e}\), \(\Sigma_{R_{e}}\), \(f_\mathrm{bulge}\), and \(Z_{*}\) exist, which is consistent with the Pearson correlation coefficients of $R_e$ in the left part of Figure \ref{fig:barunKS_score}. More compact disk galaxies tend to exhibit higher stellar metallicity and host more massive bulges and black holes. The disk scale length ($h_R$) and $R_e$ are commonly used to measure galaxy size \citep{erwinWhatDeterminesSizes2019}, and they show a strong correlation, with a Pearson correlation coefficient of 0.61. Barred galaxies tend to have smaller galaxy sizes compared to unbarred galaxies. This difference contrasts with the observation that barred galaxies are typically more extended \citep{sanchez-janssenEvidenceSecularEvolution2013,erwinWhatDeterminesSizes2019}. A more detailed discussion can be found in Section~\ref{sec:concentrated}. It is also well established that bars can influence star formation, as measured by SFR, sSFR, and $f_\mathrm{gas}$ in this context. This might explain the correlation in relatively less massive galaxies. However, this correlation becomes weak in galaxies with log \((M_\mathrm{*,30kpc}/M_\odot) >10.8\), which is out of the scope of this paper. 

There is no significant trend in the $f_\mathrm{baryon}$ between barred and unbarred galaxies. As for dark matter halo concentration, barred galaxies typically exhibit higher concentrations compared to unbarred galaxies, consistent with previous studies \citep{rosas-guevaraEvolutionBarredGalaxy2022a}. However, the difference is not statistically significant, as shown by $c_{200}$ in Figure~\ref{fig:barunKS_score}.

It is clear that it is mainly nurture that determines whether bars are present in galaxies with log \((M_\mathrm{*,30kpc}/M_\odot) >10.8\), while both nature and nurture are important in relatively less massive cases. Therefore, we divide the galaxy samples into two subgroups to conduct a more thorough investigation. In Figure \ref{fig:sixdistribu}, we compare the distribution of barred (blue), unbarred (red), and short-bar galaxies (green). The short-bar galaxies exhibit similar properties to their unbarred counterparts.

\section{Nurture in massive disk galaxies: mergers suppress or destroy bars}
\label{sec:moremassive}

Nurture, including mergers and close tidal interactions, plays a crucial role in massive galaxies. We therefore compare the properties and evolutionary history of barred, unbarred, and short-bar galaxies in the mass range log $(M_\mathrm{*,30kpc}/M_\odot) = 10.8-11.1$ in this section. 

\subsection{Massive barred galaxies have been less perturbed by mergers}
\label{sec:merger}

\begin{figure}[t]
	\centering
	\captionsetup[subfigure]{labelformat=empty}
	\subfloat[]{\includegraphics[width=\hsize]{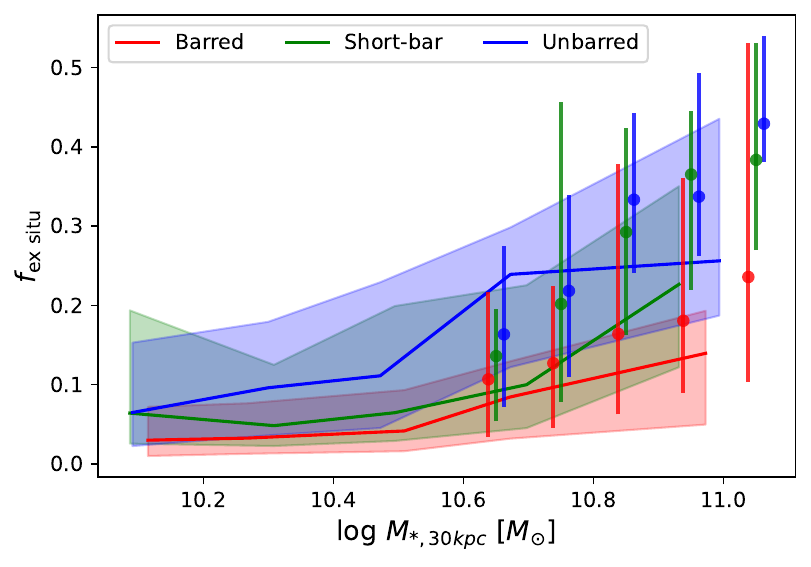}} \vspace{-20pt}
	\caption{Variation of $f_\mathrm{ex\ situ}$ with the mass of the barred (red), short-bar (green), and unbarred (blue) galaxies at $z=0$ in TNG50 (colored region) and TNG100 (points with error bars). The solid lines and the points represent the median values whose colored regions and error bars correspond to the 16th to 84th percentile.}
	\label{fig:TNG50fexsitu}
\end{figure}

\begin{figure}[t]
	\centering
	\includegraphics[width=\hsize]{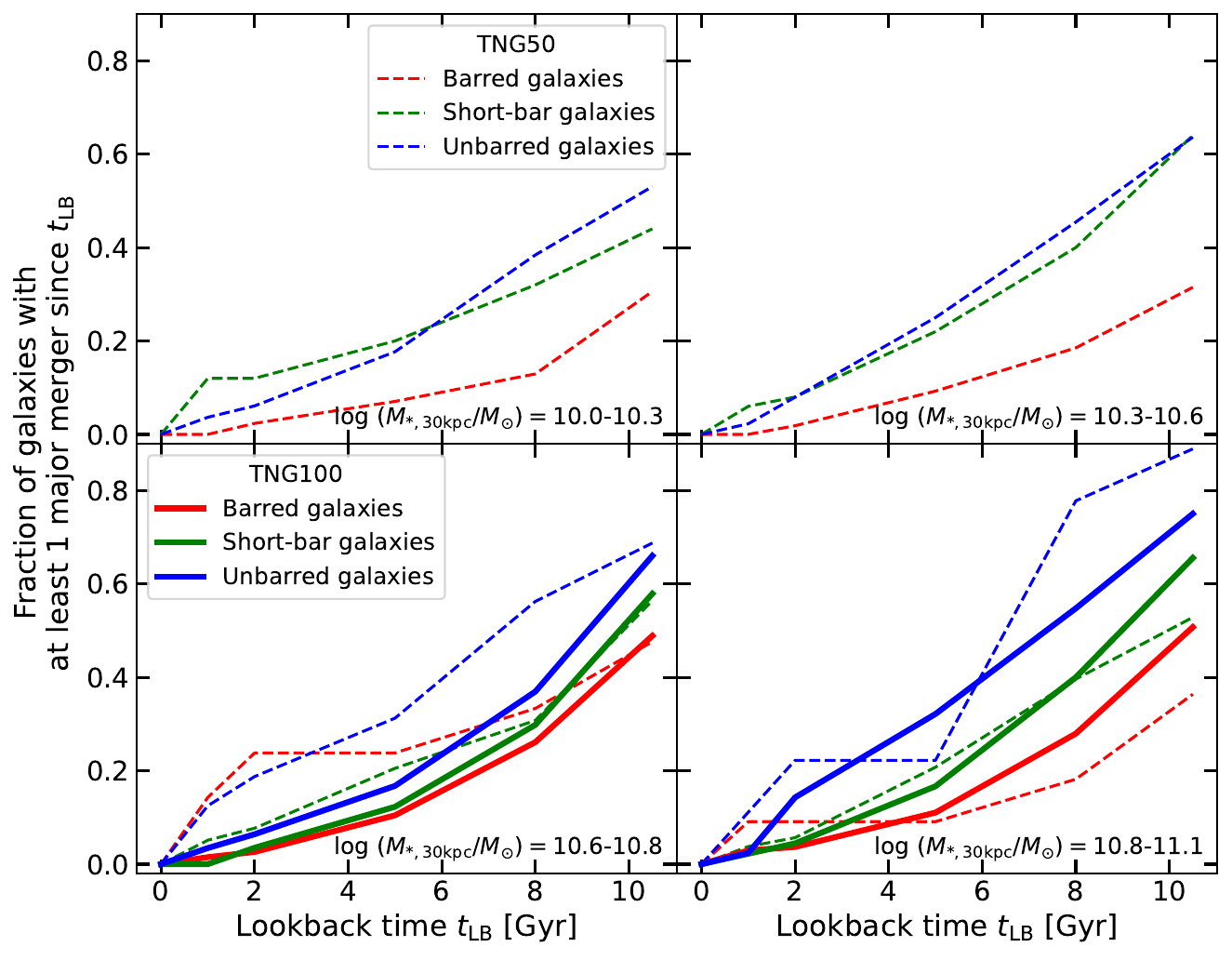}
	\caption{Fraction of galaxies that experienced at least one major merger since a certain look-back time $t_{\rm LB}$ in TNG50 (dotted) and TNG100 (solid). The results of four mass ranges are shown, i.e., log $(M_\mathrm{*,30kpc}/M_\odot)=10.6-10.8$ and log $(M_\mathrm{*,30kpc}/M_\odot)=10.8-11.1$. The blue, red, and green profiles are unbarred, barred, and short-bar galaxies, respectively.}
	\label{fig:majorminormergerframass}
 
\end{figure}

\begin{figure*}
\centering
	\captionsetup[subfigure]{labelformat=empty}
	\subfloat[]{\includegraphics[width=0.31\textwidth]{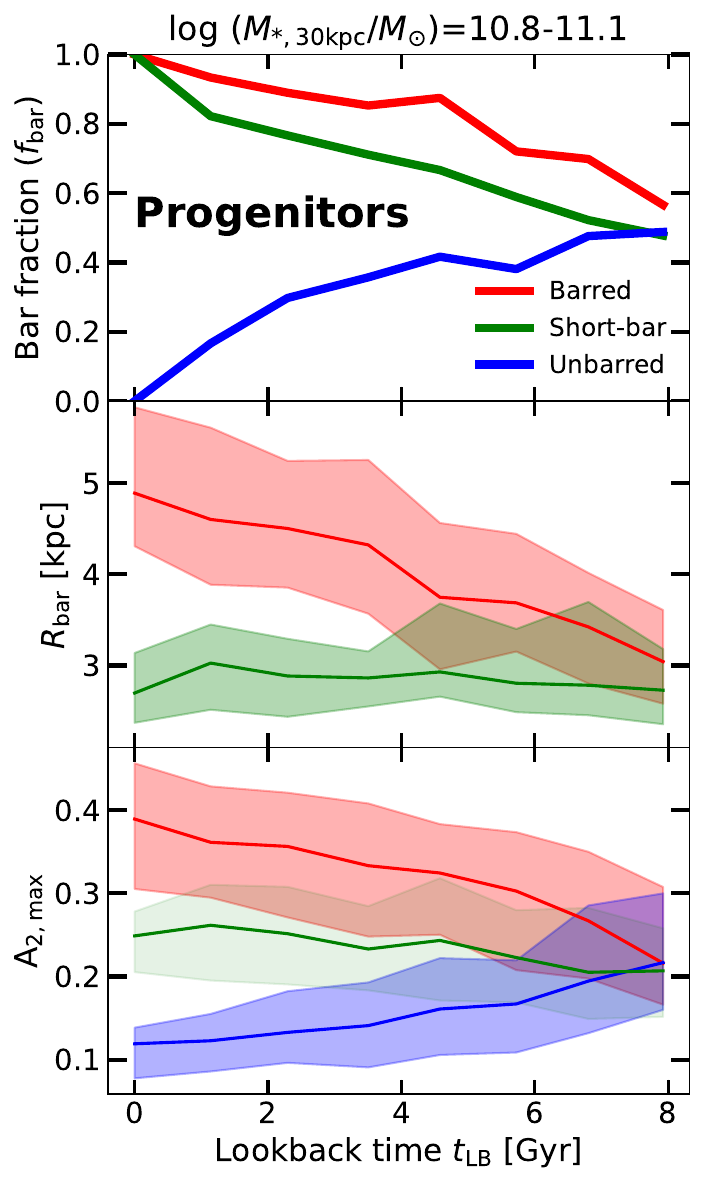}} \hspace{5pt}
 	\subfloat[]{\includegraphics[width=0.315\textwidth]{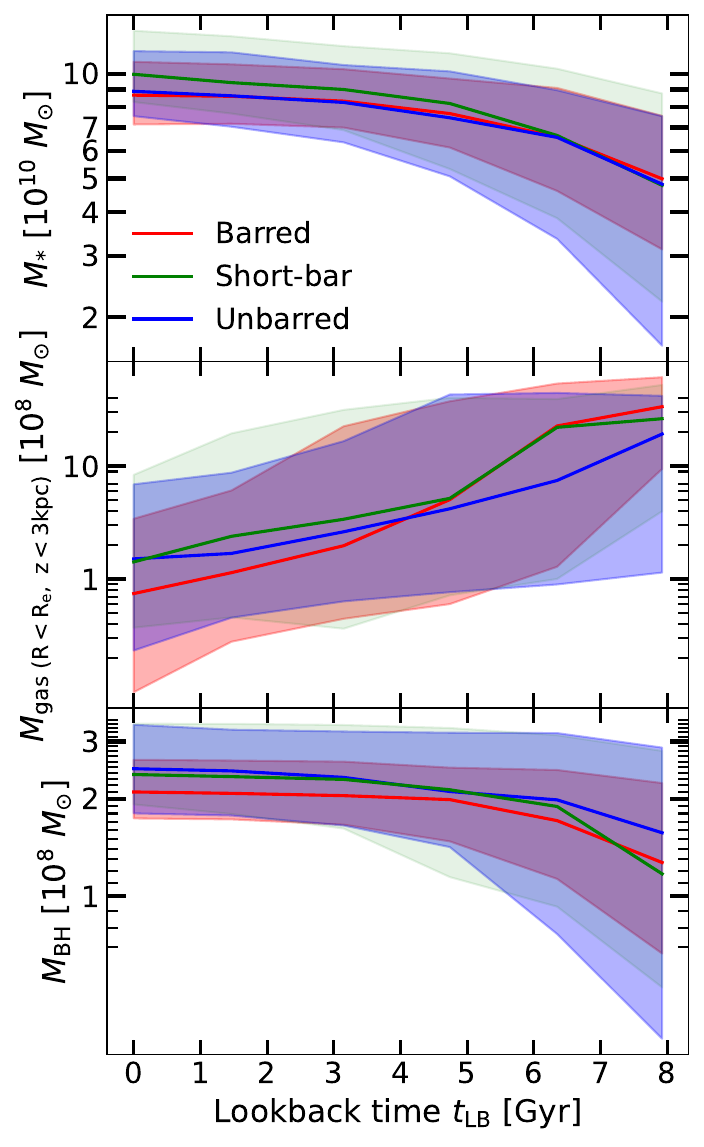}} 
	\subfloat[]{\includegraphics[width=0.315\textwidth]{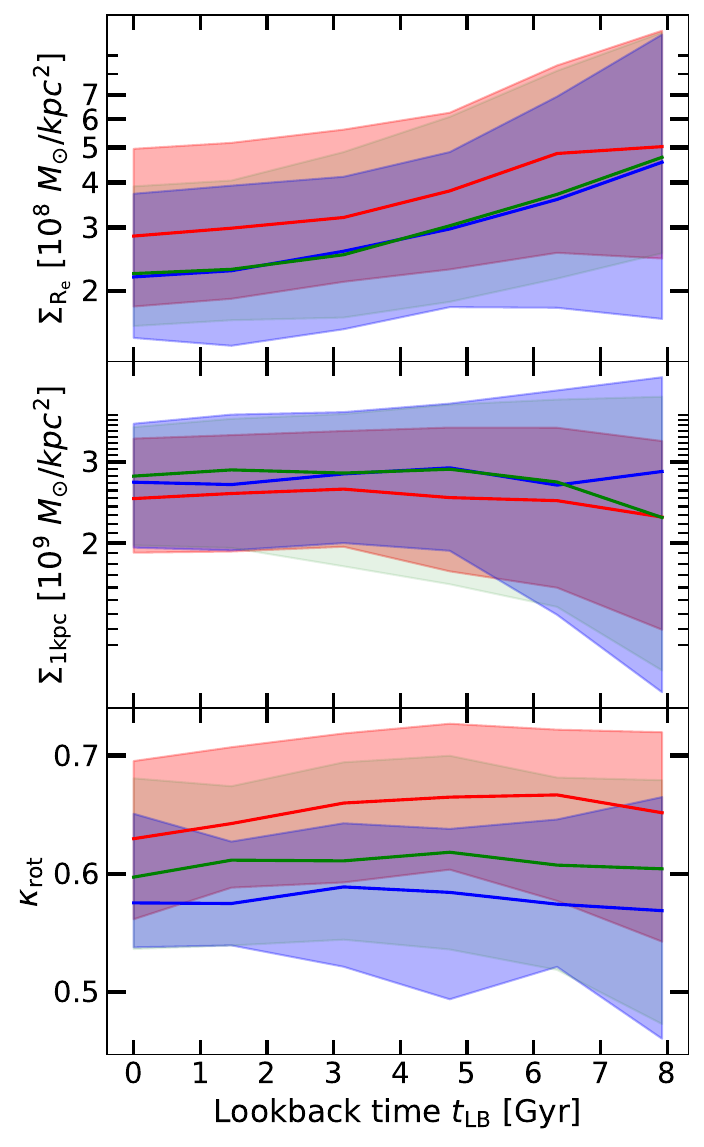}}
        \hspace{20pt}
	\caption{Evolution of the bar characteristics (left column) in massive barred (red), short-bar (green), and unbarred (blue) galaxies with log $(M_\mathrm{*,30kpc}/M_\odot)=10.8-11.1$ in TNG100 and the evolution of some parameters of the galaxies (middle and right columns). The left column displays the evolution of $R_\mathrm{bar}$ obtained from ellipse fitting (middle panel) and the evolution of bar fraction for the sample of galaxies (top panel). It also presents the evolution of $A_\mathrm{2,max}$ obtained from Fourier decomposition (bottom panel). The middle column shows the evolution of the galaxy's stellar mass ($M_{*}$), central gas mass ($M_\mathrm{gas}$), and supermassive black hole mass ($M_\mathrm{BH}$). The right column shows the evolution of the stellar surface mass density within $R_e$ ($\Sigma_\mathrm{R_{e}}$),  the central 1 kpc stellar surface mass density ($\Sigma_\mathrm{1kpc}$) and the ratio of kinetic energy in ordered rotation ($\kappa_\mathrm{rot}$). For all the panels, the x-axis is lookback time, which equals $z=0$ when lookback time = 0, and $z=1$ when lookback time $\approx$ 8. The solid lines represent the median values of each distribution, with colored regions indicating the 16th to 84th percentile range.}
	\label{fig:massiveevo}
\end{figure*}

Barred galaxies have generally experienced a weaker influence of mergers than unbarred and short-bar galaxies. As seen in the lower panels of Figure~\ref{fig:sixdistribu}, barred galaxies (red) have significantly smaller $f_{\rm halo}$ than unbarred (blue) and short-bar (green) galaxies with log $(M_\mathrm{*,30kpc}/M_{\odot})=10.8-11.1$. The median $f_\mathrm{halo}$ of barred galaxies reaches $\sim 0.15$, whereas that of unbarred galaxies is $\sim 0.25$. $f_{\rm halo}$ serves as a measure of the impact of external processes like mergers and close tidal interactions on the galaxy. The differences in $f_\mathrm{halo}$ between barred and unbarred galaxies show that bars are more likely to occur in galaxies that have undergone fewer interactions and mergers. We also examine the distribution of $f_\mathrm{ex\ situ}$ both in TNG100 and TNG50 samples. Figure~\ref{fig:TNG50fexsitu} demonstrates that in TNG50 (colored regions), the disparity in $f_\mathrm{ex\ situ}$ between barred galaxies and unbarred galaxies also increases with stellar mass. It is clear that nurture strongly influences the presence of bars.

We further examined the merger history of the galaxy samples. Figure~\ref{fig:majorminormergerframass} illustrates the major merger history of the selected sample. The bottom-right panel shows that a lower percentage of barred galaxies have experienced a major merger compared to unbarred galaxies in TNG100. Specifically, around 25\% of barred galaxies (red) have undergone major mergers during the past 8 Gyr ($t_\mathrm{LB} <8$ Gyr), while unbarred galaxies (blue) have a significantly higher rate of approximately 60\%. This significant difference in major merger history indicates that, for massive galaxies, bars are more commonly found in those with fewer mergers. This is likely because the dynamic heating from mergers can destroy existing bars or prevent their formation. The lower rate of major mergers in barred galaxies supports the idea that a quieter merger history favors the formation and maintenance of bars in massive galaxies.

\subsection{Evolutionary histories of massive barred and unbarred galaxies}
\label{sec:moreevolution}

The presence of bars in massive galaxies is primarily determined by the evolution of galaxies during $t_\mathrm{LB} = 0-8$ Gyr. The left panels of Figure~\ref{fig:massiveevo} depicts the evolution of bar fraction $f_\mathrm{bar}$, bar size $R_\mathrm{bar}$, and bar strength $A_\mathrm{2,max}$. Using the ellipse fitting method and bar identification criteria described in Section~\ref{sec:barcharacter}, the top-left panel illustrates the changes of bar fraction for the galaxy samples selected at $z=0$ during their evolution. 
The left column of Figure~\ref{fig:massiveevo} shows that both barred and unbarred galaxies exhibit a similar bar fraction of around 0.5 at $t_\mathrm{LB} = 8$ Gyr ($z\sim$1). The bar strength, represented by $\epsilon_\mathrm{max}$ and $A_\mathrm{2,max}$, is also comparable at $t_\mathrm{LB} = 8$ Gyr ($z\sim$1). This outcome implies that the progenitors of all galaxies have similar bar properties initially. However, external processes gradually erode the bars in unbarred galaxies during their subsequent evolution. In the barred sample, the bar strength, quantified by $A_\mathrm{2,max}$, and the bar size ($R_\mathrm{bar}$) increase significantly in the past 8 Gyr $(z<1)$.
A similar phenomenon was reported by \citet{athanassoulaCanBarsBe2005} using purely $N$-body simulations.

Barred galaxies have relatively higher rotation than their unbarred and short-bar counterparts as shown in the bottom-right panel of Figure~\ref{fig:massiveevo}, possibly because they have been weakly affected by mergers. Early $N$-body simulations suggested that bars tend to form rapidly once a dynamically cold disk has settled down \citep[see the review of][]{sellwoodSPIRALINSTABILITIESNBODY2012}. The difference in \(\kappa_\mathrm{rot}\) at \(t_\mathrm{LB}=8\) Gyr between barred and unbarred galaxies also suggests that bars are more likely to form in dynamically cooler galaxies. It is noteworthy that \(\kappa_{\rm rot}\) decreases around the same time that the strength of the bar increases, possibly because bars expel angular momentum to the dark matter on resonances \citep{athanassoulaBarsHaloesTheir2005}, as well as contribute to the development of boxy/peanut-shaped bulges \citep{athanassoulaBoxyPeanutBulges2016,Anderson2024}.

Barred (red), short-bar (green), and unbarred (blue) galaxies exhibit similar evolution in their central densities as measured by $\Sigma_{R_\mathrm{e}}$ and $\Sigma_\mathrm{1kpc}$, shown in Figure~\ref{fig:massiveevo}. Moreover, the differences in gas and black hole masses are also small when comparing $M_{\mathrm{gas}\ (R<R_{e},\ z<\mathrm{3kpc)}}$ and $M_\mathrm{BH}$. It is notable that there is a significantly steeper decline in gas mass within the central regions of barred galaxies compared to unbarred ones. This could be attributed to bars funneling gas into the center of galaxies, resulting in a rapid depletion of gas \citep[e.g.,][]{cheungGALAXYZOOOBSERVING2013,gavazziHa3HaImaging2015, linSDSSIVMaNGAIndispensable2020}. 

Galaxies with short bars (green in Figure~\ref{fig:massiveevo}) are likely those whose disks are moderately influenced by mergers. Their bars remain short and weak because their disks are comparatively warmer than those in barred galaxies. Moreover, the external forces they experience are not sufficient to fully destroy these short bars. Consequently, the strength of the bars remains nearly constant. In conclusion, the presence of a bar in more massive galaxies at redshift 0 is primarily determined by the strength of external influence during $t_\mathrm{LB}=0-8$ Gyr. Longer and stronger bars tend to exist in dynamically cooler disk galaxies that are less affected by mergers. See more discussions about the dynamical temperature in Section \ref{sec:toomre-q}.

\section{Nature in less massive disk galaxies: barred galaxies tend to follow the compact evolutionary pathway}
\label{sec:lessmassive}

The difference in merger history between barred and unbarred galaxies with stellar masses log $(M_*/M_\odot)=10.6-10.8$ is relatively minor, as shown in Figures \ref{fig:TNG50fexsitu} and \ref{fig:majorminormergerframass}. This distinction becomes even smaller for galaxies with log $(M_*/M_\odot)<10.6$ in the TNG50 simulation (Figure \ref{fig:TNG50fexsitu}). The spatial resolution of TNG50 is likely adequate to reasonably resolve bars in these less massive disk galaxies. The bar fraction in TNG50 of this mass range roughly agrees with observations of disk galaxies (red dotted profile in Figure \ref{fig:TNG50vsTNG100vsS4Gselect}). It is clear that factors such as conditions at $z>1$ and internal dynamical processes play crucial roles in determining whether a galaxy develops a bar.

Therefore, this section focuses on the nature of less massive barred galaxies in TNG50 and TNG100, by examining their size and evolutionary history in detail.

\subsection{Barred galaxies are more compact than unbarred cases}
\label{sec:concentrated}

\begin{figure}[t]
	\centering
	\captionsetup[subfigure]{labelformat=empty}
	\subfloat[]{\includegraphics[width=\hsize]{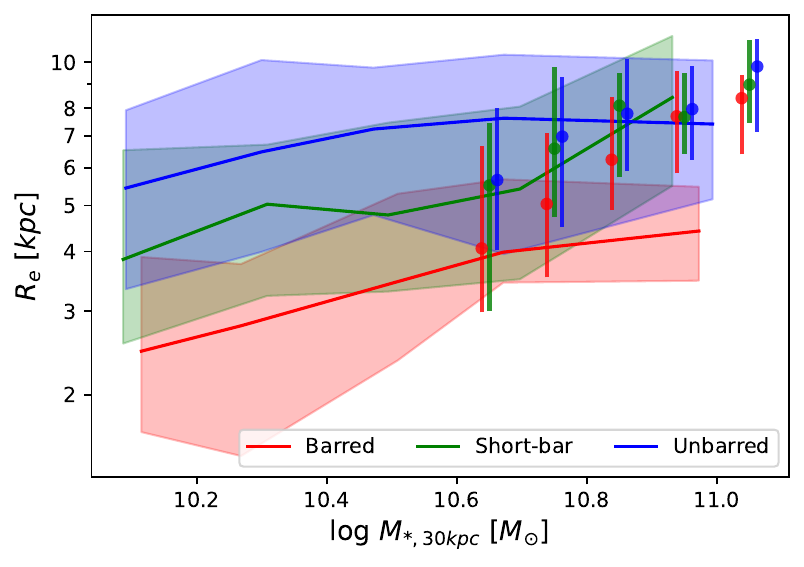}} \vspace{-20pt}
	\caption{Variation of $R_\mathrm{e}$ with the mass of the barred (red), short-bar (green), and unbarred (blue) galaxies at $z=0$ in TNG50 (colored regions) and TNG100 (points with error bars). The solid lines and the points represent the median values, with colored regions and error bars indicating the range from the 16th to 84th percentile of each distribution.}
	\label{fig:TNG50Re}
\end{figure}

\begin{figure*}
\centering
	\captionsetup[subfigure]{labelformat=empty}
	\subfloat[]{\includegraphics[width=0.31\textwidth]{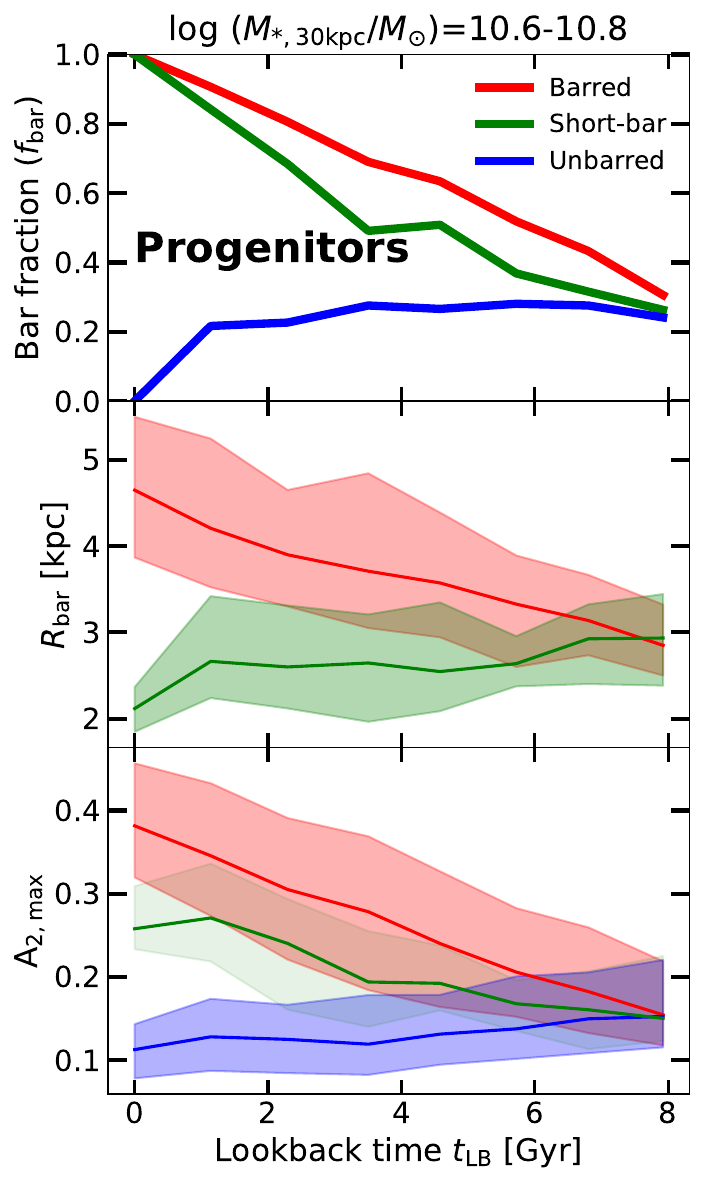}} \hspace{5pt}
 	\subfloat[]{\includegraphics[width=0.315\textwidth]{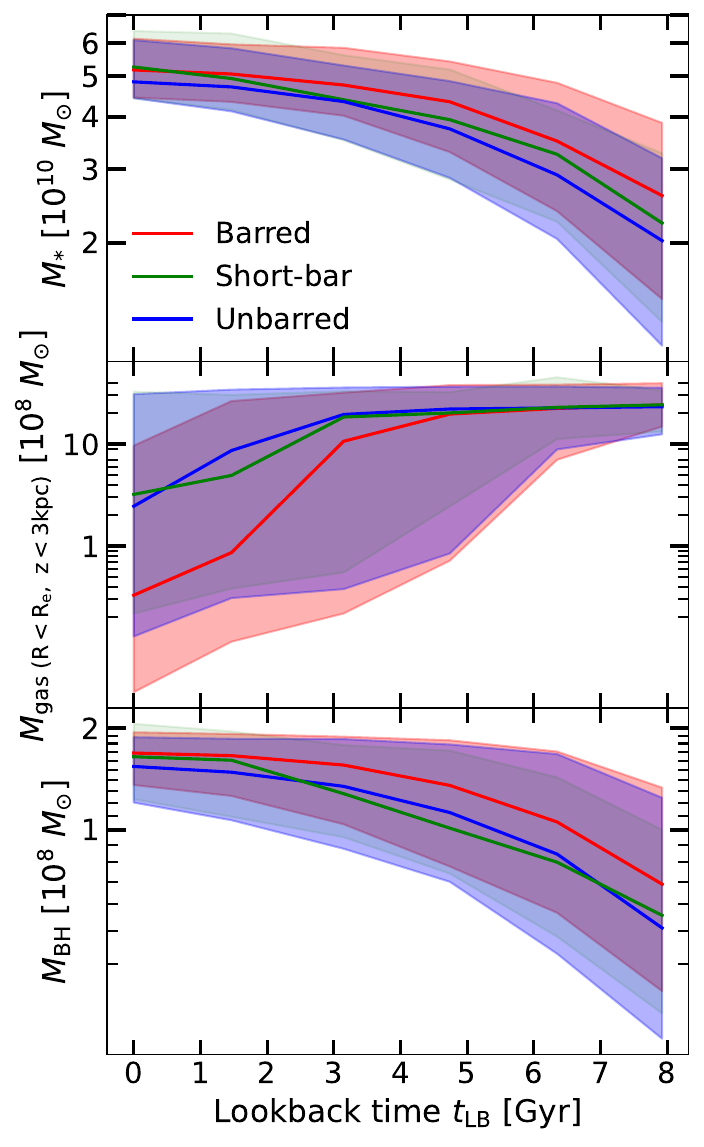}} 
	\subfloat[]{\includegraphics[width=0.3223\textwidth]{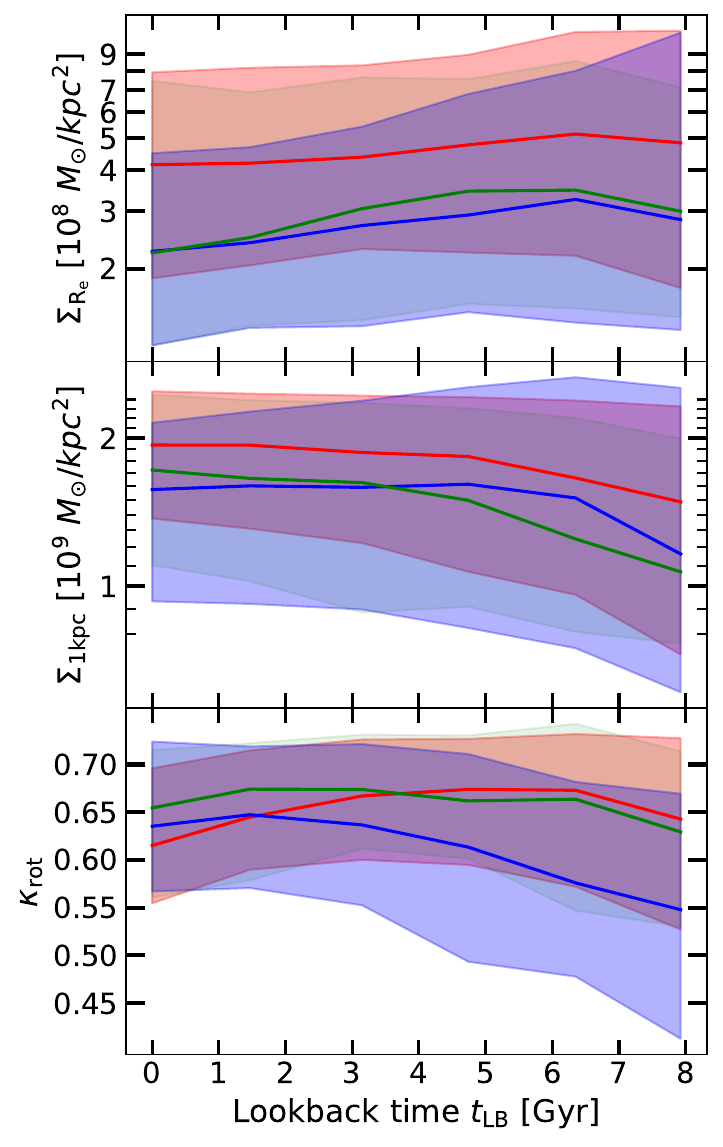}}
	\caption{Evolution of the bar characteristics (left column) in relatively less massive barred (red), short-bar (green), and unbarred (blue) galaxies with log $(M_\mathrm{*,30kpc}/M_\odot)=10.6-10.8$ in TNG100 and the evolution of some parameters of the galaxies (middle and right columns). The left column displays the evolution of the bar size ($R_\mathrm{bar}$) obtained from ellipse fitting (middle panel) and the evolution of bar fraction for the sample of galaxies based (top panel). It also presents the evolution of $A_\mathrm{2,max}$ obtained from Fourier decomposition (bottom panel). The middle column shows the evolution of stellar mass ($M_{*}$), central gas mass ($M_\mathrm{gas}$), and supermassive black hole mass ($M_\mathrm{BH}$). The right column shows the evolution of the stellar surface mass density within $R_e$ ($\Sigma_\mathrm{R_{e}}$), the central 1 kpc stellar surface mass density ($\Sigma_\mathrm{1kpc}$) and the ratio of kinetic energy over ordered rotation ($\kappa_\mathrm{rot}$). For all the panels, the $x$-axis is lookback time, which equals $z=0$ when lookback time = 0 Gyr, and $z=1$ when lookback time $\approx$ 8 Gyr. The solid lines represent the median values of each distribution, with colored regions indicating the 16th to 84th percentile range.}
	\label{fig:lessmassiveevo}
\end{figure*}

Barred galaxies tend to be more compact than unbarred galaxies in TNG100. As shown in Figure~\ref{fig:sixdistribu}, barred galaxies generally have smaller $R_{e}$ values than unbarred galaxies, particularly for less massive galaxies with log $(M_\mathrm{*,30kpc}/M_\odot)=10.6-10.8$, meaning that barred galaxies are more compact. 

We use data from the TNG50 simulation to explore the influence of compactness further. Figure~\ref{fig:TNG50Re} shows the relationship between $R_{e}$ and log $M_\mathrm{*,30kpc}$ for barred and unbarred galaxies in TNG50 (colored region) and TNG100 (points with error bars). There is a significant difference in $R_{e}$ at lower masses between barred and unbarred galaxies. Most barred galaxies have $R_{e}< 4$ kpc, while unbarred galaxies generally have $R_{e}$ larger than 4 kpc, in agreement with TNG100. As presented in \citet{duOriginRelationStellar2022} and \citet{duPhysicalOriginMassSize2024}, the sizes of disk galaxies in TNG simulations are primarily determined by the specific angular momenta inherited from their parent dark matter halos. Most disk galaxies we select here have a halo fraction $f_\mathrm{halo}$ less than 0.2, which means they align with the ``mass-size-angular momentum-X'' scaling relation proposed by \citet{duPhysicalOriginMassSize2024} and \citet{maEvolutionaryPathwaysDisk2024}, where ``X'' can be the metallicity, age, and central density of galaxies. In this picture, the diverse mass-size relations observed in galaxies are largely determined by the angular momentum of their host galaxies \citep[e.g.,][]{Fall&Efstathiou1980, Mo1998}. Extended galaxies then are unbarred because they are weakly responsive to bar instabilities. This suggests that the paucity of barred galaxies in less massive galaxies may be caused by the more extended nature of many of these disk galaxies. Our findings suggest that the presence of both barred and unbarred galaxies can be explained, to some extent, by the mass-size relation. It is also worth noting that no clear correlation between bars and environment has been found \citep{dengDependenceGalacticBars2023}. 

It is also important to mention that, in observations, barred galaxies typically have larger sizes compared to unbarred galaxies. As shown in Figure 3 of \citet{sanchez-janssenEvidenceSecularEvolution2013}, the disk scale length of barred galaxies is 15\% larger than that of unbarred galaxies. Similarly, in Figure 9 of \citet{erwinWhatDeterminesSizes2019}, across a wide range of stellar masses (\(10^9 - 10^{11} M_\odot\)), barred galaxies are slightly more extended than their unbarred counterparts. This is somehow inconsistent with our results from TNG, where barred galaxies tend to be more compact than unbarred galaxies, as seen in TNG50, TNG100, and other simulations like EAGLE \citep{algorryBarredGalaxiesEAGLE2017}. The origin of this discrepancy between simulations and observations is unclear; exploring this discrepancy further is out of the scope of this paper. Here, we merely note it exists.

\begin{figure*}[ht!] 	
\centering 	
\includegraphics[width=\textwidth]{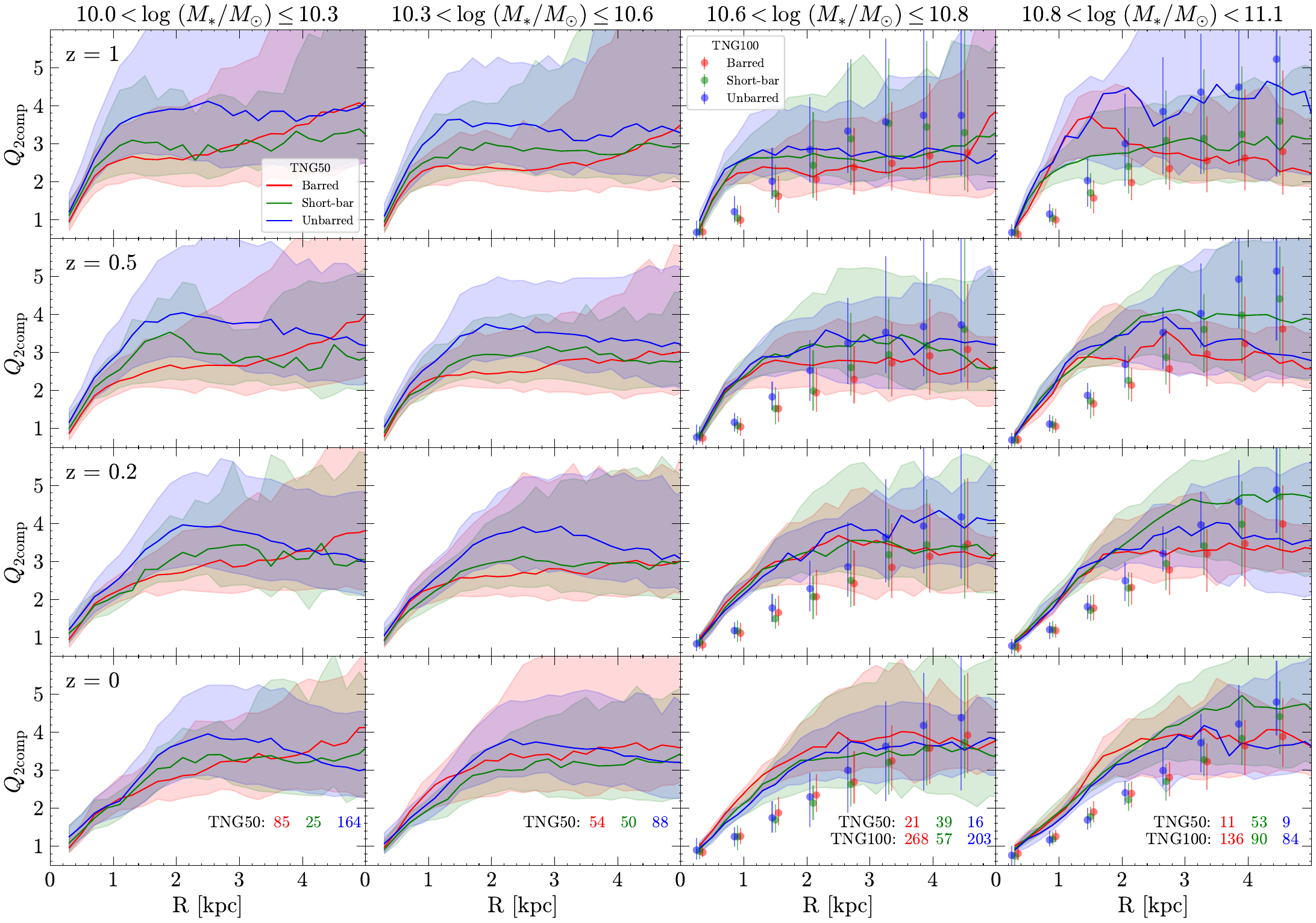}\vspace{-10pt} 	
\caption{Evolution of the radial profiles of two-component disk stability parameter ($Q_{\rm 2comp}$) for barred (red), short-bar (green), and unbarred (blue) galaxies across four mass ranges from $z=1$ (top) to $z=0$ (bottom) in TNG50 (colored region). The columns from left to right, show the changes over cosmic time within four mass ranges shown at the top of each column. TNG100 galaxies (points with error bars) are overlaid in the mass bins with $\log\ (M_{*}/M_{\odot})=10.6-10.8$ and $\log\ (M_{*}/M_{\odot})=10.8-11.1$ (right two columns). The solid curves and points represent the median values, while the colored regions and error bars give the range from the 16th to the 84th percentiles of each $Q_{\rm 2comp}$ distribution. The numbers in the bottom panels are the number count of galaxies in each sample.} 	\label{fig:ToomreQTNG100TNG50} 	 
\end{figure*}

\subsection{Barred galaxies follow a compact evolutionary pathway}
\label{sec:lessevolution}

The left column of Figure~\ref{fig:lessmassiveevo} illustrates the temporal evolution of bar fraction (top), bar size (middle), and $A_\mathrm{2,max}$ (bottom) for the less massive sample. Consistent with the situation found in more massive galaxies, the fraction and strength of bars in barred, short-bar, and unbarred galaxies are nearly identical at $t_\mathrm{LB} = 8$ Gyr.

We trace the evolution of each galaxy component and find that barred galaxies assemble their stellar masses earlier, which is consistent with the evolution of compact disk galaxies selected in \citet{maEvolutionaryPathwaysDisk2024}.
The middle column of Figure~\ref{fig:lessmassiveevo}
shows the temporal evolution of $M_\mathrm{{*}}$, $M_\mathrm{gas\ (R<R_{e},\ z<3kpc)}$, and $M_\mathrm{BH}$ for barred, unbarred, and short-bar galaxies. At a lookback time of $t_\mathrm{LB}$ = 8 Gyr (z$\sim$1), as illustrated in the top panel, barred galaxies have approximately 30\% higher galaxy stellar mass than unbarred galaxies, indicating they assembled earlier. 

High mass density and galaxy rotation are more likely to lead to the formation of bars. The right column in Figure~\ref{fig:lessmassiveevo} shows the evolution of $\Sigma_\mathrm{R_e}$ (top), $\Sigma_\mathrm{1kpc}$ (middle), and $\kappa_\mathrm{rot}$ (bottom) in less massive galaxies with log $(M_\mathrm{*,30kpc}/M_\odot)=10.6-10.8$. Since $t_\mathrm{LB}$ = 8 Gyr, $\Sigma_\mathrm{R_e}$ and $\Sigma_\mathrm{1kpc}$ of barred galaxies consistently exceed those of unbarred galaxies throughout this time period. This suggests that bars tend to form in galaxies with a higher stellar density, and they can maintain a high density during their evolution. Furthermore, barred galaxies tend to exhibit a higher $\kappa_\mathrm{rot}$ during the initial phases at $t_\mathrm{LB}=4$ Gyr. The decline in $\kappa_\mathrm{rot}$ for barred galaxies can be ascribed to the onset of buckling instability or angular momentum redistribution or the scarcity of gas necessary for the creation of rapidly rotating new stars.

Unexpectedly, unbarred galaxies show a consistent rise in $\kappa_{\rm rot}$ (as shown in the bottom-right panel of Figure~\ref{fig:lessmassiveevo}) toward the present, reaching high rotation rates by $z=0$. This suggests that for these extended galaxies, even with such a high $\kappa_{\rm rot}$, bar formation is inefficient. The large size and high gas fraction of these disks may contribute to their stability against bar instabilities.

In summary, relatively compact galaxies with older stellar populations are more likely to develop bars, following the compact evolutionary pathway \citep{Du_2021, maEvolutionaryPathwaysDisk2024}. While bars can influence a galaxy's shape, star formation, and kinematics over the long term, they are unlikely to alter the overall evolutionary pathways of galaxies.

\subsection{Evidence of quenching induced by bars?}
\label{sec:sfr}

Barred galaxies exhibit somewhat lower sSFR compared to unbarred galaxies at \( z=0 \). The KS test in Figure~\ref{fig:barunKS_score} indicates that barred galaxies typically have lower sSFR than their unbarred counterparts, particularly in less massive disk galaxies ($\log (M_\mathrm{*,30kpc}/M_\odot) = 10.6 - 10.8$). This difference is minimal in more massive galaxies. The histograms in the upper-right panel of Figure~\ref{fig:sixdistribu} clearly demonstrate that barred galaxies are more likely to be quenched. Bars are thought to suppress star formation through several mechanisms. One such mechanism involves the shocks and shears created by a bar, which can prevent gas from collapsing and thus inhibit star formation within most of the bar \citep{khoperskovBarQuenchingGasrich2018}. Another mechanism is that the internal gas flow driven by bars can hasten the consumption of cold gas \citep[e.g.,][]{robertsGasDynamicsBarred1979,athanassoulaExistenceShapesDust1992a,spinosoBardrivenEvolutionQuenching2017,linSDSSIVMaNGAIndispensable2020}.  Figure \ref{fig:lessmassiveevo} shows the evolution of gas mass ($M_\mathrm{gas}$) in barred galaxies (red colored region in the middle panel). It is clear that gas in barred galaxies depletes faster than in both unbarred (blue) and short-bar (green) cases. However, the more compact nature of barred galaxies leads to a faster gas consumption rate during their early evolutionary stages, as suggested by \citet{maEvolutionaryPathwaysDisk2024}. Therefore, we cannot entirely dismiss the possibility that barred galaxies quench more rapidly because they exhaust their gas reserves earlier through accelerated evolution of the entire galaxy rather than gas inflows induced by bars. It is likely that all these mechanisms are occurring simultaneously.

More expected is the lack of a difference in the sSFR between barred and unbarred galaxies within the mass range \(\log (M_\mathrm{*,30kpc}/M_\odot) = 10.8 - 11.1\). This implies that the mere existence of bars is insufficient to quench star formation in massive disk galaxies. Consequently, the concurrent presence of bars and reduced sSFR observed might originate from a shared underlying factor rather than a direct cause-and-effect relationship. Nonetheless, an increased frequency of mergers could also stimulate star formation in massive unbarred galaxies, potentially aligning their sSFR with that of barred galaxies by $z=0$. Once more, given that all these processes are unfolding concurrently, their individual significance remains unclear.

\section{Unbarred galaxies are more extended and dynamically hotter}
\label{sec:toomre-q}

The Toomre-$Q$ parameter is often used to quantify the dynamical temperature of a stellar disk \citep{toomreGravitationalStabilityDisk1964}. In addition to stars, the galaxy disk also contains a gas component with different kinematics, which contributes a low velocity dispersion component that maintains the dynamical responsiveness of the combined disk of stars and gas \citep{sellwoodSecularEvolutionDisk2014}.
We analyze such a two-component disk using the approximate formula of \citet{romeoEffectiveStabilityParameter2011}, which is an improved form of the\citet{wangGravitationalInstabilityDisk1994} approximation:

\begin{equation}
    \frac{1}{Q_{\rm 2comp}}=
    \left\{ \begin{aligned}        
    & \frac{W}{Q_{*}}+\frac{1}{Q_{\rm g}} & (Q_{*}>Q_{\rm g}),   \\   
    & \frac{1}{Q_{*}}+\frac{W}{Q_{\rm g}} & (Q_{*}<Q_{\rm g}), \\       
    \end{aligned} \right.
\end{equation}

\begin{equation}
    W=\frac{2\sigma_* \sigma_{\rm g}}{\sigma_*^2+\sigma_{\rm g}^2},
\end{equation}
where $Q_{*}=\sigma_* \kappa/ \pi G\Sigma_{*}$ and $Q_{\rm g}=\sigma_{\rm g} \kappa/ \pi G\Sigma_{\rm g}$ are the stellar and gaseous Toomre parameters. Here, $\kappa$, $\Sigma_*$, $\Sigma_g$, $\sigma_*$, $\sigma_{\rm g}$, and $G$ represent the epicyclic frequency, the stellar surface density, the gaseous surface density, the stellar radial velocity dispersion, the gaseous radial velocity dispersion, and the gravitational constant, respectively. The gas and stars within $|z|<3$ kpc are used to calculate $\Sigma$ and $\sigma$. Figure~\ref{fig:ToomreQTNG100TNG50} shows the radial distribution of $Q_{\rm 2comp}$ for two-component disks for barred (red), short-bar (green), and unbarred galaxies (blue) across four mass ranges from $z=1$ (top) to $z=0$ (bottom) in TNG50. Galaxies in TNG100 are overlaid in the mass bins with $\log\ (M_*/M_{\odot})=10.6-10.8$ and $\log\ (M_*/M_{\odot})=10.8-11.1$ (right two columns).

Barred galaxies typically have lower $Q_{\rm 2comp}$ compared to unbarred galaxies at early evolutionary stages ($z=1$). At $z=1$, barred 
galaxies in both TNG50 (colored regions) and TNG100 (points with error bars) show systematically lower $Q_{\rm 2comp}$ than unbarred galaxies within the radial range of $1<R<5$ kpc, with a more significant difference in the lower mass range. This is consistent with previous studies showing that disks with lower Toomre-$Q$ tend to form bars more quickly \citep{athanassoulaBisymmetricInstabilitiesKuz1986,sahaWhyAreGalaxies2018}. \citet{Du2015} showed, using N-body simulations, that if the initial Toomre-$Q$ of a disk is greater than 2.2, then the disk is unresponsive to bar instabilities. Similarly here, at $z=1$, barred galaxies have $Q_\mathrm{2comp}$ $\sim$ 2.6 while unbarred galaxies have $Q_\mathrm{2comp}$ $\gtrsim$ 3. \citet{fragkoudiBarFormationEvolution2024} also found that barred galaxies have lower Toomre-$Q$ compared to unbarred galaxies from the zoom-in Auriga simulations. We verify the correlation between bar formation and Toomre-$Q$ in this study covering a mass range of log $(M_\mathrm{*,30kpc}/M_\odot)=10.6-11.1$ in TNG100 and log $(M_\mathrm{*,30kpc}/M_\odot)=10-11.1$ in TNG50.

The elevated $Q_\mathrm{2comp}$ values in unbarred galaxies with a lower stellar mass during the early stages are likely attributable to their disks not being fully settled. As explained in Sections~\ref{sec:moremassive} and \ref{sec:lessmassive}, such unbarred galaxies form later than their barred counterparts, thus having larger $Q_\mathrm{2comp}$ due to their shallow potential well. This is different from the cases with a higher stellar mass where mergers are responsible for the high $Q_\mathrm{2comp}$ in unbarred galaxies, as illustrated in Figure~\ref{fig:majorminormergerframass}.

\begin{figure}
	\centering
	\includegraphics[width=1\columnwidth]{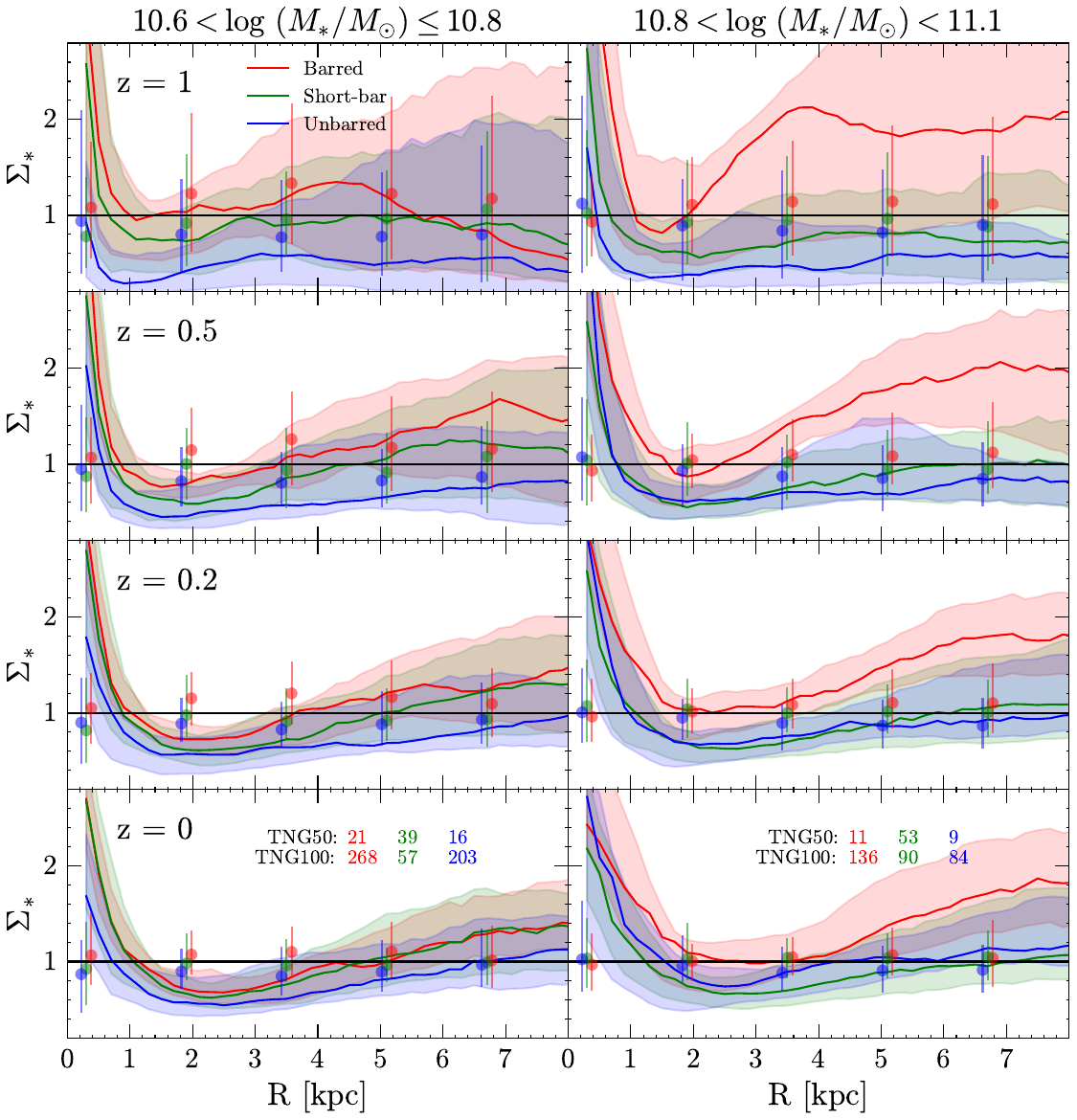}

	\caption{Radial profiles of stellar surface density for barred (red), short-bar (green), and unbarred (blue) galaxies in TNG50 (shaded areas) and TNG100 (points with error bars) from $z=1$ (top row) to $z=0$ (bottom row). In each panel, all distributions are rescaled by the median radial stellar surface density of the disk galaxy sample from TNG100. The black horizontal line, which corresponds to 1, represents the median radial stellar surface density profile of the TNG100 disk galaxy sample. Points and lines represent the median values of each distribution, while the error bars and shaded areas show the 16th to 84th percentiles.}
	\label{fig:Sigma_star_dif}
\end{figure}

Within the mass range of log $(M_\mathrm{*,30kpc}/M_\odot)=10.6-11.1$, the comparison of the radial distribution of $Q_\mathrm{2comp}$ for barred, short-bar, and unbarred galaxies across TNG50 and TNG100 shows that disk galaxies in TNG50 generally display higher $Q_\mathrm{2comp}$ values at radii less than $3$ kpc compared to their TNG100 counterparts, with the largest discrepancy reaching approximately $1$. This elevated $Q_\mathrm{2comp}$ likely accounts for the tendency of bars to be shorter in TNG50 within the log $(M_\mathrm{*,30kpc}/M_\odot)=10.6-11.1$ mass range, as depicted in Figure~\ref{fig:TNG50vsTNG100vsS4Gselect}.

The stellar mass distribution in the central parts of galaxies is strikingly different, perhaps due to the different gravitational resolutions employed in TNG50 and TNG100. This largely leads to the deviation of Toomre-$Q$ between TNG50 and TNG100 in the central regions, shown in Figure \ref{fig:ToomreQTNG100TNG50}. In Figure~\ref{fig:Sigma_star_dif}, we present the radial profiles of stellar surface density in TNG50 compared to that in TNG100 for barred, short-bar, and unbarred galaxies. From an early epoch ($z = 1$) to $z = 0$, galaxies in TNG50 display significantly higher stellar surface densities within 1 kpc than those in TNG100, while the surface density of such galaxies at $R = 1 - 3$ kpc is 50\% lower. The lower stellar density in the 1-3 kpc range in TNG50 results in a higher Toomre-$Q$ in this region compared to TNG100. \citet{frankelSimulatedBarsMay2022a} also found that TNG50-2, which has a gravitational softening length twice that of TNG50-1, exhibits systematically longer bars. They speculated that the differences in softening length influence the mass distribution, potentially affecting the properties of the bars.  Previous N-body simulations have similarly shown that smaller gravitational softening lengths tend to result in shorter bars \citep{bauerCanStellarDiscs2019}. Moreover, the higher central mass density can change the dynamical structure of the bar \citep[e.g.,][]{pfennigerDissipationBarredGalaxies1990,shenDestructionBarsCentral2004a}. This outcome indicates that a sufficiently high resolution is necessary for accurately reproducing the dynamic process of galaxies, particularly in the central regions. Such a difference in bars then can lead to a more significant difference over the whole galaxy.

\section{Formation and evolution of bars and their sizes}
\label{sec:barlength}

\begin{figure*}
\centering
        \captionsetup[subfigure]{labelformat=empty}
	\subfloat[]{\includegraphics[width=0.95\textwidth]{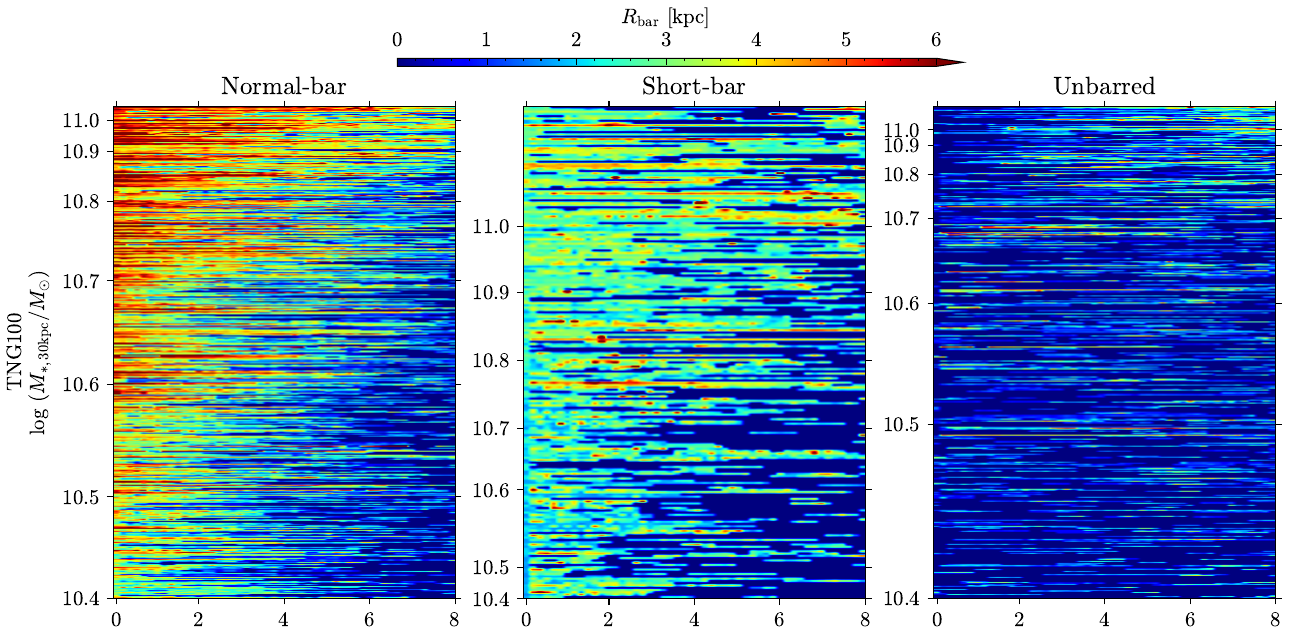}} \vspace{-15pt}
 	\subfloat[]{\includegraphics[width=0.95\textwidth]{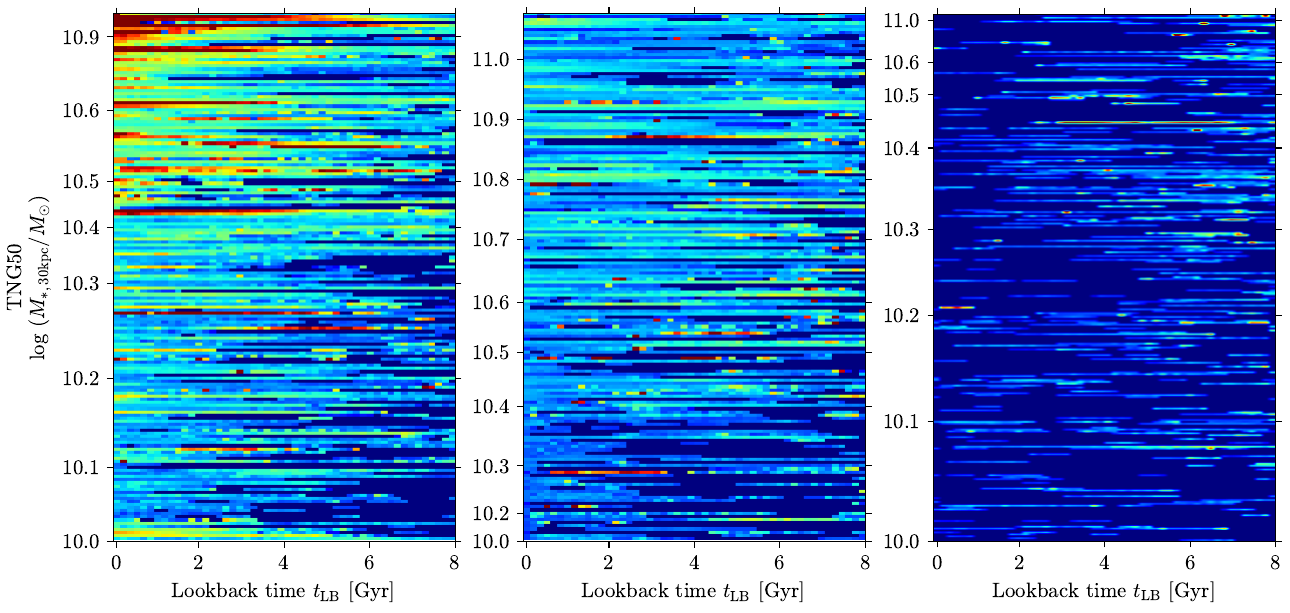}} 

	\caption{Evolutionary history of bar size (color) for barred (left column), short-bar (middle column), and unbarred (right column) galaxies in TNG100 (top row) and TNG50 (bottom row). Each row of one panel represents the bar size of an individual galaxy as a function of lookback time ($t_\mathrm{LB}$). Galaxies in each panel are ranked by stellar mass.}
	\label{fig:TNG100barlengthevo}
\end{figure*}

\begin{figure}
	\centering
	\includegraphics[width=0.9\columnwidth]{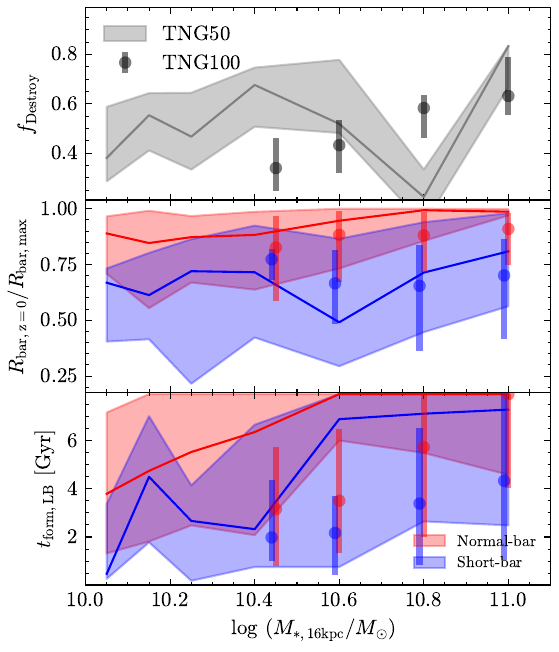}
	\caption{Evolutionary features of bars as a function of galaxy stellar mass. TNG100 galaxies are shown by points with error bars. And the shaded regions correspond to TNG50 galaxies. The top panel shows the ratio of unbarred galaxies $f_\mathrm{Destroy}$ at $z=0$ that previously had a bar. The middle panel shows the ratio of the bar size between $z = 0$ and its maximum size throughout the galaxy's history $R_{\mathrm{bar, z = 0}} / R_{\mathrm{bar, max}}$. The bottom panel illustrates the bar formation time in lookback time $t_\mathrm{form,LB}$ as a function of galaxy mass. For the middle and bottom panels, the points and solid lines represent the median values, and the error bars and shaded envelopes show the 16th to 84th percentile distributions. Regarding the bottom panel, for galaxies with masses greater than $10^{10.6}\, M_{\odot}$, many bars had already formed by a lookback time of 8 Gyr.  In the top panel, the error bars and shaded regions stand for variations resulting from different definitions of bar presence.}
	\label{fig:diversebarmode}
\end{figure}

In the local Universe, bar sizes $R_{\rm bar}$ vary from $\sim$100s pc to nearly 10 kpc \citep[e.g.,][]{hoyleGalaxyZooBar2011,diaz-garciaCharacterizationGalacticBars2016,erwinDependenceBarFrequency2018}. The size of these bars is a fundamental factor that influences their host galaxies, with longer bars having an impact on larger areas. Therefore, it is important to also study how the bar size changes.

We trace the presence and evolution of bars and their sizes in the parent sample of disk galaxies selected at $z=0$, corresponding to those shown in Figure~\ref{fig:TNG50vsTNG100vsS4Gselect}. These include disk galaxies at $z=0$ with log $(M_\mathrm{*,30kpc}/M_\odot)=10.4-11.1$ in TNG100 and log $(M_\mathrm{*,30kpc}/M_\odot)=10.0-11.1$ in TNG50. Figure~\ref{fig:TNG100barlengthevo} presents the evolution of bar size since $t_\mathrm{LB} = 8$ Gyr for galaxies classified as barred, short-bar, and unbarred at $z = 0$. Each row in every panel represents a single galaxy. In the early stages, irregular structures may be misidentified as bars. Thus, we ignore bar signals that are continuously shorter than 4 snapshots ($\sim0.5$ Gyr) as transient features. It is clear that bars tend to form earlier in more massive galaxies, namely ``downsizing'' \citep{Anderson2024}. During $t_\mathrm{LB}=4-8$ Gyr, more massive galaxies are generally more likely to have a bar than their lower-mass counterparts. This tendency is clear across all disk galaxies in both TNG50 and TNG100, as depicted in Figure~\ref{fig:TNG100barlengthevo}. The bottom panel of Figure~\ref{fig:diversebarmode} shows the age of bars as a function of stellar mass for both TNG50 (shaded regions) and TNG100 (points with error bars) galaxies. Bar ages of both normal (red) and short (blue) cases increase with the galaxy's stellar mass. This behavior has also been reported in the formation of bars in TNG50 by \citet{Anderson2024}, which is in line with observational results \citep{shethEvolutionBarFraction2008,cuomoRelationsStructuralParameters2020}. Moreover, we come across cases of bar ``renewal", in which bars reform or are rejuvenated within a galaxy. This phenomenon has been noted in previously isolated galaxy simulations \citep{sellwoodFormationDiskGalaxies1999,bournaudGasAccretionSpiral2002,combesPatternSpeedEvolution2008}, perhaps because of the existence of newly formed, dynamically responsive stellar components.

As depicted in the top panel of Figure~\ref{fig:diversebarmode}, normal-sized bars typically keep growing longer and stronger, which is also shown in Figures~\ref{fig:massiveevo} and ~\ref{fig:lessmassiveevo}. These bars are likely to extend to their co-rotation radii, i.e., known as ``fast bars'' \citet{debattistaConstraintsDynamicalFriction2000}. By contrast, a large proportion of short bars become shorter, thus forming ``slow bars''. This result accounts for why many galaxies are ``slow", as presented in \citet{frankelSimulatedBarsMay2022a}.

\subsection{Unbarred galaxies: bars are either destroyed or never formed}
\label{sec:barlengthofbarunbar}

A significant number of unbarred galaxies have experienced bar formation at some stage, although a considerable fraction have never developed a bar throughout their history. As shown in the right column of Figure~\ref{fig:TNG100barlengthevo}, a large proportion of galaxies have at some point developed a bar, with some bars persisting for a long time. These long-lived bars are potentially the result of the global bar instability in especially massive galaxies, though they may be eradicated through violent mergers. A substantial fraction of these galaxies form short-lived bars that quickly dissipate. These transient bars are likely provoked by close tidal interactions with neighboring galaxies, resulting in the absence of stable bar orbits. 

In the top panel of Figure~\ref{fig:diversebarmode}, we show the fraction of unbarred galaxies that have had bars in the past in TNG50 (solid lines with shaded regions) and TNG100 (points with error bars) as a function of galaxy mass. The solid lines and points stand for galaxies where bar signals are continuously detected for at least 8 snapshots (around 1 Gyr). Meanwhile, the shaded regions and error bars represent the upper and lower limits based on the definitions of bars lasting at least 4 and 12 snapshots respectively. For TNG100, an obvious trend can be seen: as the galaxy mass decreases, the proportion of unbarred galaxies that have never formed a bar rises. This phenomenon is weak in TNG50, which is likely due to the weaker numerical issues in TNG50 because of its higher resolution. Generally speaking, unbarred galaxies can be divided into two groups: those whose bars have been destroyed and those that have never had bars, with each group making up about 50\% of the unbarred population. 

\subsection{Short bars: either born short or become short}
\label{sec:barlengthshort}

Massive galaxies with stellar masses exceeding $10^{10.6} M_{\odot}$ in the TNG100 simulation exhibit a bar fraction that is approximately 10\%–20\% higher than what is observed, which can be attributed to excess production of short bars. This overabundance of short bars is even more pronounced in the TNG50 simulation, as depicted in Figure \ref{fig:TNG50vsTNG100vsS4Gselect}. We traced the formation time of bars at $z=0$, denoted as $t_\mathrm{form,LB}$, which corresponds to the lookback time when the bar first meets the bar criteria described in Section~\ref{sec:barcharacter}. Short bars are generally old structures, although some of them seem to form slightly later than normal bars by approximately 1 - 2 Gyr, as shown in the bottom panel of Figure~\ref{fig:diversebarmode}.

Most short-bar galaxies appear to form their bars early, in a way similar to galaxies with standard-sized bars, as shown in the middle column of Figure~\ref{fig:TNG100barlengthevo}. Short bars can be formed either by a reduction in length or by having low elongation. The ratio $R_\mathrm{bar, z=0} / R_\mathrm{bar, max}$ serves as a metric to gauge the extent of bar size changes. $R_{\mathrm{bar, max}}$ is estimated as the average of the three largest bar sizes during the galaxy's evolution, and $R_{\mathrm{bar, z = 0}}$ is the average bar size measured from the three most recent snapshots. Approximately 28\% of bars in short-bar galaxies shrink to less than half of their maximum length, such that $R_{\mathrm{bar, z = 0}}/R_{\mathrm{bar, max}}<0.5$. In comparison, 23\% of these bars have not undergone significant shrinkage or elongation, with a ratio $R_{\mathrm{bar, z = 0}}/R_{\mathrm{bar, max}}>0.8$.

The shortening of bars may be attributed to merger events. Short-bar galaxies exhibit higher \(f_\mathrm{ex\ situ}\) (Figure~\ref{fig:TNG50fexsitu}) and \(f_\mathrm{halo}\) (Figure~\ref{fig:sixdistribu}) values. It is well known that mergers can weaken or even destroy bars by heating the outer parts of galactic disks \citep[e.g.,][]{ghoshFateStellarBars2021}, thus giving rise to unbarred disk galaxies. Short bars tend to persist, as they are less vulnerable to destruction in the case that mergers are not powerful enough to impact the central zones of galaxies. As presented in Sections \ref{sec:moremassive} and \ref{sec:lessmassive}, short-bar galaxies fall between barred and unbarred galaxies in terms of their merger histories. Therefore, a higher prevalence of short bars in TNG simulations indicates that its merging processes may be less efficient in destroying bars than those in the real Universe.

Another possible explanation for the abundance of short bars in TNG simulations is that the galaxies in these simulations are less centrally concentrated. Short-bar galaxies tend to have larger effective radii (\(R_e\)), indicating that these galaxies are less compact and their mass is less concentrated, as shown in Figure~\ref{fig:TNG50Re}. Figure~\ref{fig:TNG50Re} illustrates that galaxies with short bars are larger in size than those with bars having normal sizes. TNG simulations do not fully capture the evolution of the central regions of galaxies. The large softening length is likely to create a weaker central concentration, which may lead to a higher frequency of short bars. 

Other subgrid physical processes may alse influence the size of bars. As shown in the center panels of Figure~\ref{fig:massiveevo} and Figure~\ref{fig:lessmassiveevo}, short bars typically contain more gas in their inner regions compared to normal bars. This gas might prevent these bars from slowing down by transferring angular momentum to the bars. \citet{semczukPatternSpeedEvolution2024} investigated the evolution of bar pattern speed in TNG50 and found that AGN feedback could play a role in removing the gas from the central regions of barred galaxies, thereby facilitating the slowdown and elongation of bars. Similarly, \citet{rosas-guevaraGalaxyFormationPhysics2024} found that feedback processes, including AGN feedback and supernova (SN) feedback, can also impact the presence and size of bars by using zoom-in cosmological simulations.

Furthermore, these short bars may evolve into the secondary or inner bars of double-barred galaxies when a primary or outer bar independently forms via bar instabilities \citep{Du2015}. Indeed, Integral Field Unit (IFU) observations have detected fairly old secondary bars \citep{2019MNRAS.484.5296D}. Although the TNG simulations have achieved remarkable success, short bars are still difficult to be fully understood. Nevertheless, TNG indicates that short bars are likely to be commonly present in galaxies. They may play a vital role in the evolution of galaxies, particularly in the growth of bulges \citep{guoNewChannelBulge2020} and supermassive black holes \citep[e.g.,][]{shlosmanBarsBarsMechanism1989, duBlackHoleGrowth2017, liHowNestedBars2023} in the central regions of galaxies. We recommend that future studies exclude these short-bar galaxies from the general barred galaxies in TNG100, considering the low resolution of recent cosmological simulations and the complex formation mechanisms of short bars, when they are only concerned with the general characteristics of barred galaxies. These short-bar galaxies are more similar to unbarred galaxies than to barred galaxies in many galaxy properties, as detailed in Appendix~\ref{sec:shorbar_ks}.

\section{Conclusions}
\label{sec:conclusion}

In this study, we have performed a statistical analysis to explore the properties of barred galaxies using the cosmological simulations IllustrisTNG-100 and -50. We selected samples of 417 barred galaxies, 307 unbarred disk galaxies, and 114 short-bar galaxies at $z=0$ from TNG100 to investigate the differences in their formation. Additionally, all conclusions are validated using TNG50.

We apply KS tests on various galaxy parameters comparing barred and unbarred galaxies at $z=0$ to investigate the factors contributing to the formation of bars in disk galaxies. The sample is divided into less massive (log $(M_\mathrm{*,30kpc}/M_\odot)<10.8$) and more massive (log $(M_\mathrm{*,30kpc}/M_\odot)>10.8$) bins. We examine the evolution of bar properties in the sample using ellipse fitting and Fourier decomposition techniques over a lookback time of $t_\mathrm{LB}$ = 0-8 Gyr ($z=0-1$). Additionally, we also study the evolution of both the Toomre-$Q$. Our main findings are summarized as follows.
\begin{enumerate}[wide,label=\arabic*.]
  \item The presence of a bar is mainly determined by the evolution of disk galaxies over $t_\mathrm{LB}$ = 0-8 Gyr. Bars tend to form earlier in more massive galaxies, namely ``downsizing''. At $t_\mathrm{LB}$ = 8 Gyr ($z=1$), the bar strength and bar fractions in barred, unbarred, or short-bar are almost identical (see Figure~\ref{fig:massiveevo} and Figure~\ref{fig:lessmassiveevo}). 
  \item Nurture suppresses or destroys bars in massive disk galaxies with log $(M_\mathrm{*,30kpc}/M_\odot)>10.8$. Barred galaxies undergo significantly fewer mergers compared to unbarred ones (Figure~\ref{fig:majorminormergerframass}). As a result, they have a smaller mass fraction of (kinematically-derived) stellar halos and ex-situ stars. Most unbarred galaxies once host a bar (Figure~\ref{fig:TNG100barlengthevo}). A more violent assembly history leads to a dynamically hotter disk, which results in unbarred galaxies.
  \item Both nature and nurture play an important role in determining the properties of bars in galaxies with $10.6< \log (M_\mathrm{*,30kpc}/M_\odot)<10.8$. In this mass range, Barred galaxies are relatively compact, evolving along the compact evolutionary pathway. Their extended morphology leads to dynamically hotter disks, thus becoming or remaining unbarred.
  \item Short bars are limited in growth or shortened (Figure~\ref{fig:TNG100barlengthevo}). Galaxies with short bars have experienced only moderate effects from mergers. Central mass densities are the second-most influential factor affecting the size of bars. Galaxies with shorter bars typically have lower \SigmaRe.
  \item Barred disk galaxies at $z=0$ show a higher percentage of quenched galaxies compared to unbarred galaxies. This can be partly explained by either a faster depletion of gas in barred galaxies or following a more compact evolutionary pathway in nature.
  \item Galaxies with short bars are more similar to unbarred galaxies. Therefore, these galaxies should either be excluded from the analysis of barred galaxies or classified into the sample of unbarred galaxies.
  
\end{enumerate}

Our results provide insights into the properties and evolution of barred galaxies in the cosmological simulation IllustrisTNG. This study highlights the complex interplay between internal factors, such as stellar mass distribution, and external factors, such as galaxy mergers, in determining the size of bars and the evolution of barred galaxies of different masses. Further research is needed to delve deeper into the specific mechanisms and processes involved in the evolution of bars and their relationship with the internal and external environments of galaxies.

\begin{acknowledgements}

The authors acknowledge the support by the Fundamental Research Funds for the Central Universities (No. 20720230015), the Natural Science Foundation of Xiamen China (No. 3502Z202372006), the Science Fund for Creative Research Groups of the National Science Foundation (NSFC) of China (No. 12221003), and the China Manned Space Program through its Space Application System. S.L. also acknowledges the support by XMU Training Program of Innovation and Enterpreneurship for Undergraduates (No. 2022Y1567). The TNG100 and TNG50 simulations used in this work, two of the flagship runs of the IllustrisTNG project, have been run on the HazelHen Cray XC40-system at the High-Performance Computing Center Stuttgart as part of project GCS-ILLU of the Gauss centers for Supercomputing (GCS). This work is also strongly supported by the Computing Center in Xi’an.  
\end{acknowledgements}

\bibliographystyle{aa}
\bibliography{ref}

\begin{appendix}
\onecolumn
\section{Measurement of Disk Scale Length}
\label{sec:parameter h}

The disk scale length is derived from the decomposition of the galaxy's stellar surface density image. We use the \texttt{Imfit} package \citep{erwinIMFITFASTFLEXIBLE2015}, to fit the image using standard $\chi^2$ minimization and the Nelder-Mead minimizer. If the bar is omitted in the fitting process, 2-component decompositions (Sérsic bulge + exponential disk) will underestimate disk scale length of barred galaxies \citep{saloSPITZERSURVEYLAR2015}. So we fit the galaxies using three components:
\begin{itemize}
    \item The profile of bulge component is given by the \citet{sersicAtlasGalaxiasAustrales1968} function:
    \begin{equation}
        \Sigma(R) = \Sigma_e{\rm exp}\left\{-b_n \left[\left(\frac{R}{R_e} \right)^{1/n}-1 \right]  \right\},
    \end{equation}
    where $\Sigma_e$ and $R_e$ represent the surface density and the radius encompassing half of the total mass of the bulge. The Sérsic-index $n$ describes the shape of the radial profile, which becomes steeper with increasing $n$. The factor $b_n$ is a normalization constant determined by $n$.
    \item The disk component is described by an infinitesimally thin exponential disk:
    \begin{equation}
        \Sigma(R) = \Sigma_0 {\rm exp} \left(-\frac{R}{h_R} \right),
    \end{equation}
    where $\Sigma_0$ is the central surface density and $h_R$ is the scale length of disk.
    \item The bar component is described by a 2D analog of the classic \citet{ferrersPotentialsEllipsoidsEllipsoidal1877} ellipsoid. The surface density function is:
    \begin{equation}
        \Sigma(m) = 
        \begin{cases}
            \Sigma_0(1-m^2)^n & {\rm if}\ m<1 \\
            0 & {\rm otherwise},
        \end{cases}
    \end{equation}
    where $\Sigma_0$ is the central region and $n$ defines the sharpness of the bar outer region, and $m$ is the elliptical radius, defined (assuming ellipticity = $1-b/a$) by
    \begin{equation}
        m^2 = \left(\frac{|x|}{a} \right)^{c_0+2}+
        \left(\frac{|y|}{b} \right)^{c_0+2}.
    \end{equation}
    Here $c_0$ controls the shape of the bar, with values of $c_0<0$ correspond to disky shape while $>0$ describe boxy shape and $0$ corresponds to a perfect ellipse.
\end{itemize}

\begin{figure}[h!]
\centering
        \captionsetup[subfigure]{labelformat=empty}
	\subfloat[]{\includegraphics[width=0.5\columnwidth]{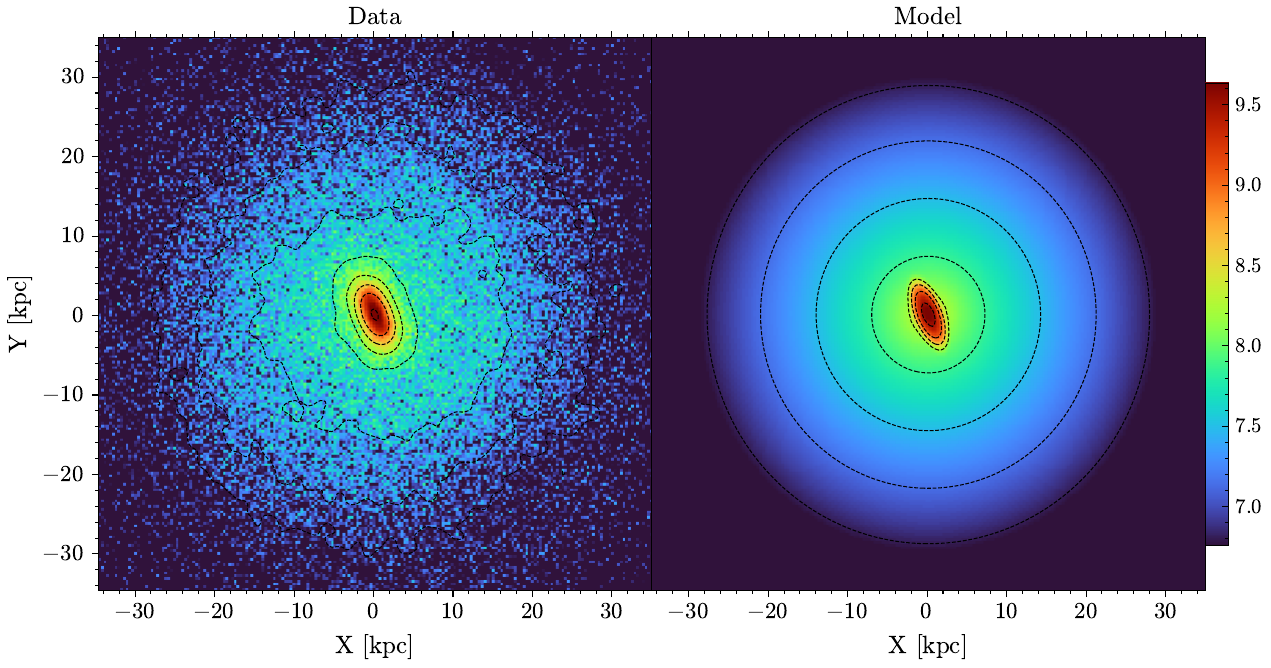}} \vspace{-15pt}
 	\subfloat[]{\includegraphics[width=0.5\columnwidth]{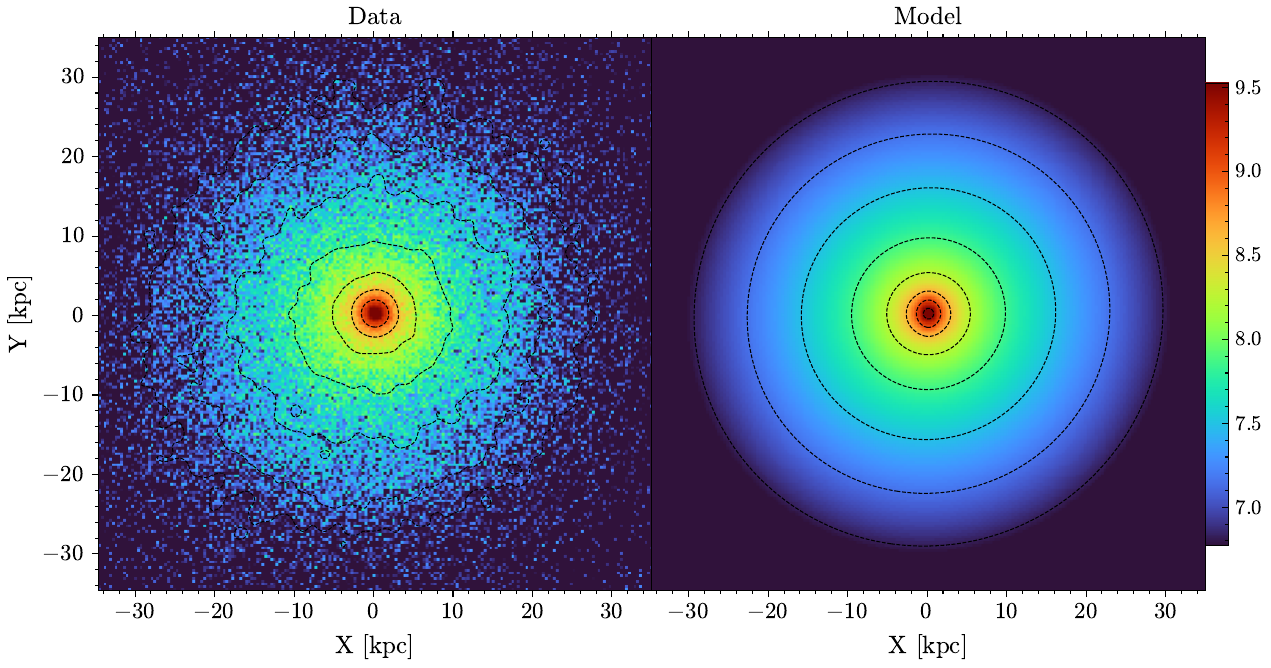}} 
    \vspace{-15pt}
	\caption{Examples of face-on stellar surface density images of a barred galaxy (top) and an unbarred (bottom) from TNG100. The right panels show the corresponding fitting results from \texttt{Imfit}.}
	\label{fig:TNG100imfit}
\end{figure}

\section{The Difference of Short-bar vs. Barred and Unbarred Galaxies}
\label{sec:shorbar_ks}
\begin{figure*}[h!]
\centering
        \captionsetup[subfigure]{labelformat=empty}
	\subfloat[]{\includegraphics[width=0.8\textwidth]{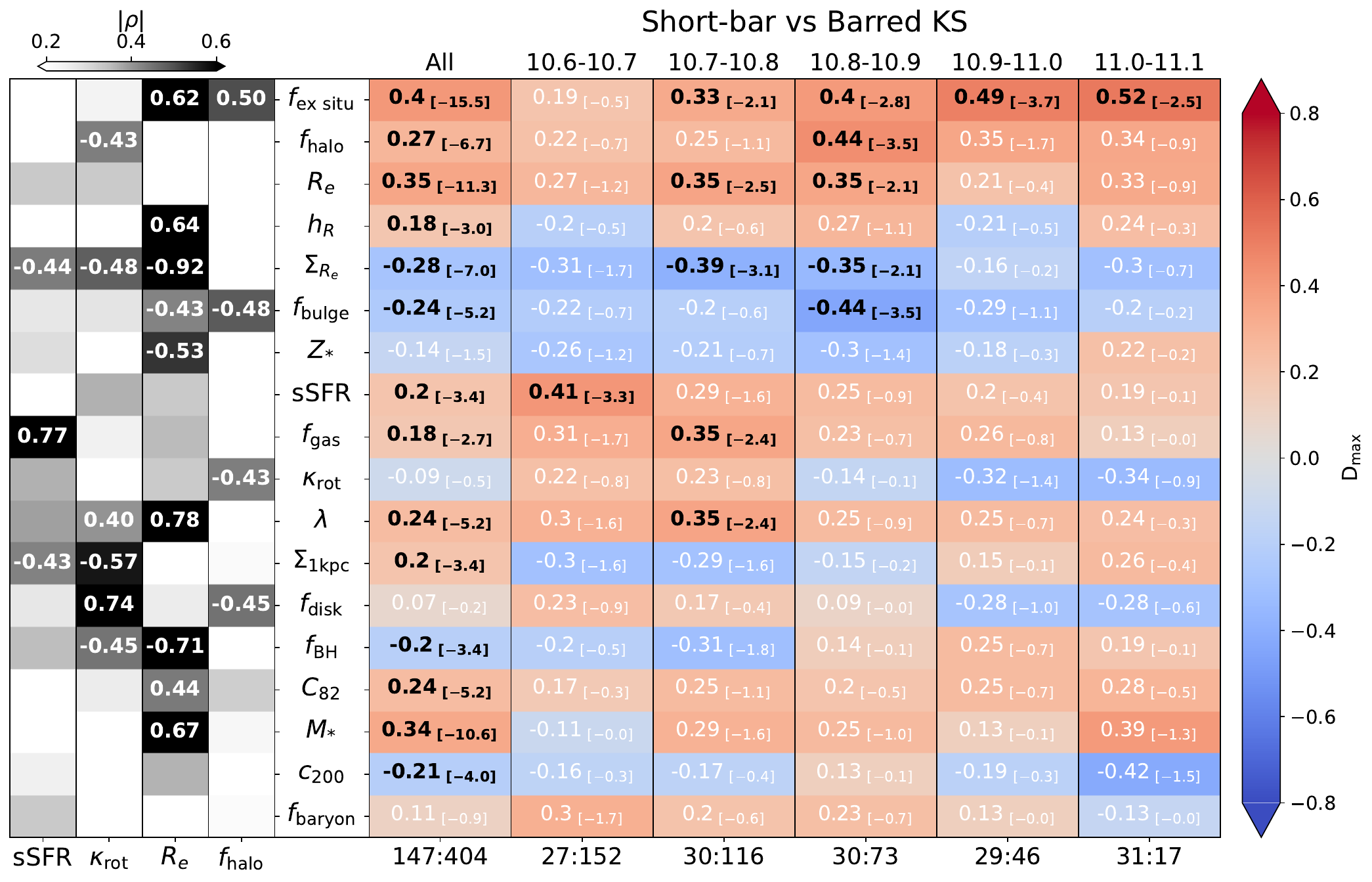}} \vspace{-15pt}
 	\subfloat[]{\includegraphics[width=0.8\textwidth]{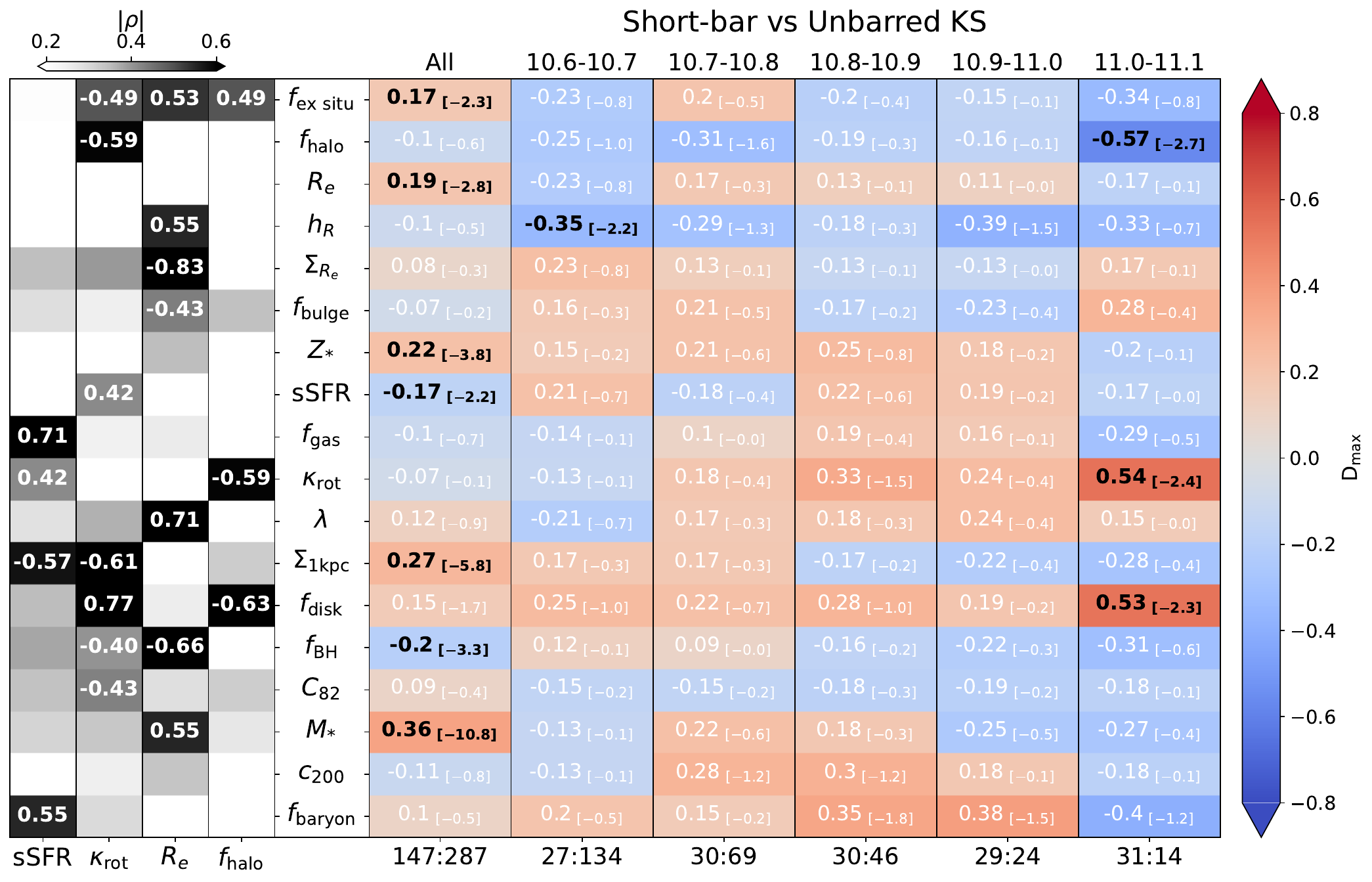}} 
    \vspace{-15pt}
	\caption{Similar to Fig~\ref{fig:barunKS_score}, this figure presents KS test results of TNG100 galaxies: the top panel shows the comparison between short-bar and barred galaxies, and the bottom panel shows results between short-bar and unbarred galaxies.}
	\label{fig:KS_short_bar}
\end{figure*}

Similar to those shown in Fig~\ref{fig:barunKS_score} and discussed in section~\ref{sec:KS}, we perform KS tests comparing short-bar galaxies with both barred galaxies and unbarred galaxies. As shown in Fig~\ref{fig:KS_short_bar}, short-bar galaxies exhibit larger differences in galaxy parameters when compared to barred galaxies than to unbarred galaxies, as indicated by the higher $|D_\mathrm{max}|$ values and the many bold values (representing p-values$<0.01$). Overall, the properties of short-bar galaxies are more similar to those of unbarred galaxies than to those of barred galaxies.

\end{appendix}

\end{document}